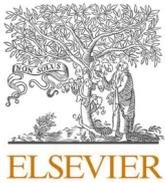
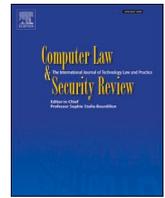

# The Fundamental Rights Impact Assessment (FRIA) in the AI Act: Roots, legal obligations and key elements for a model template

Alessandro Mantelero

*Jean Monnet Chair in Mediterranean Digital Societies and Law, Polytechnic University of Turin, Corso Duca degli Abruzzi, 24, 10129 Torino, Italy*

ABSTRACT

What is the context which gave rise to the obligation to carry out a Fundamental Rights Impact Assessment (FRIA) in the AI Act? How has assessment of the impact on fundamental rights been framed by the EU legislator in the AI Act? What methodological criteria should be followed in developing the FRIA? These are the three main research questions that this article aims to address, through both legal analysis of the relevant provisions of the AI Act and discussion of various possible models for assessment of the impact of AI on fundamental rights.

The overall objective of this article is to fill existing gaps in the theoretical and methodological elaboration of the FRIA, as outlined in the AI Act. In order to facilitate the future work of EU and national bodies and AI operators in placing this key tool for human-centric and trustworthy AI at the heart of the EU approach to AI design and development, this article outlines the main building blocks of a model template for the FRIA. While this proposal is consistent with the rationale and scope of the AI Act, it is also applicable beyond the cases listed in Article 27 and can serve as a blueprint for other national and international regulatory initiatives to ensure that AI is fully consistent with human rights.

## 1. Introduction

The latest wave of AI development, beginning in the first decade of the 21st century, has raised serious concerns about the multifaceted implications of the various technologies under the AI umbrella, despite the significant positive opportunities they bring. As with steam power, automobiles and many other technologies, the early large-scale adoption of AI has once again posed 'the gift of the evil devil' dilemma,[1] as shown by the current debate on the use of LLMs (Large Language Model) and the emergence of various critical issues,[2] from 'hallucinations',[3] to data

---

E-mail address: Alessandro.mantelero@polito.it.

[1] See Guido Calabresi, *Ideals, Beliefs, Attitudes, and the Law: Private Law Perspectives on a Public Law Problem* (Syracuse University Press 1985), Ch. 1.

[2] See also Council of Europe – European Commission for the Efficiency of Justice (CEPEJ), 'Use of Generative Artificial Intelligence (AI) by judicial professionals in a work-related context. Information note prepared by the CEPEJ Working group on Cyberjustice and Artificial Intelligence (CEPEJ-GT-CYBERJUST)' (2024), Strasbourg, CEPEJ-GT-CYBERJUST(2023)5final <https://rm.coe.int/cepej-gt-cyberjust-2023-5final-en-note-on-generative-ai/1680ae8e01> accessed 3 July 2024.

[3] So-called AI 'hallucination' is a phenomenon in which a large language model (LLM), such as a generative AI chatbot, identifies patterns or objects that do not exist and produces output that is nonsensical or completely inaccurate. On this anthropomorphisation of machine learning, which began with the same notion of AI, and its implications for the social acceptance of AI, see Nicholas Barrow, 'Anthropomorphism and AI Hype' (2024) AI and Ethics, https://doi.org/10.1007/s43681-024-00454-1; Luciano Floridi and Anna C Nobre, 'Anthropomorphising Machines and Computerising Minds: The Crosswiring of Languages between Artificial Intelligence and Brain & Cognitive Sciences' (2024) 34(5) Minds and Machines https://doi.org/10.1007/s11023-024-09670-4; James Johnson, 'Finding AI Faces in the Moon and Armies in the Clouds: Anthropomorphising Artificial Intelligence in Military Human-Machine Interactions' (2024) 38(1) Global Society 67–82, https://doi.org/10.1080/13600826.2023.2205444; Adriana Placani, 'Anthropomorphism in AI: Hype and Fallacy' (2024) AI and Ethics, https://doi.org/10.1007/s43681-024-00419-4; Amani Alabed, Ana Javornik, and Diana Gregory-Smith, 'AI anthropomorphism and its effect on users' self-congruence and self–AI integration: A theoretical framework and research agenda' (2022) 182 Technological Forecasting and Social Change 121786, https://doi.org/10.1016/j.techfore.2022.121786; Mark Coeckelbergh, 'Three Responses to Anthropomorphism in Social Robotics: Towards a Critical, Relational, and Hermeneutic Approach' (2022) 14(10) International Journal of Social Robotics 2049–2061, https://doi.org/10.1007/s12369-021-00770-0. See also Sandra Wachter et al., 'Do Large Language Models Have a Legal Duty to Tell the Truth?' (2024) <https://doi.org/10.2139/ssrn.4771884> accessed 26 June 2024.

protection[4] and copyright infringement.[5]

The introduction of a disruptive technology into society – capable of changing the paradigm of a wide range of human activities and influencing social behaviour in various way – usually generates a combination of high expectations and serious concerns. These concerns are not limited to the difficulty of changing the paradigm and facing the associated negative effects, such as on the labour market in the case of AI, but are largely related to issues affecting the early stages of any new technology.

It is not surprising that, faced with the uncertain scenario of the AI revolution and the difficulty of existing laws to provide specific solutions, the first response of policymakers was to expand regulatory systems and, in the absence of adequate provisions, to look for general ethical principles for AI. This approach led to a proliferation of ethics codes and guidelines developed by a variety of national and international public and private bodies.[6]

Although this first ethics-focused regulatory exercise led to a set of almost common core values,[7] most with a significant legal dimension, their implementation can cover a wide range of options, making this principles-based approach too vague to adequately address the challenges of AI. On the one hand, this left the AI industry a wide margin of manoeuvre in aligning ethical values with their business interests.[8] On the other hand, the general nature of these principles and their formulation, such as sustainable AI or human oversight, did not provide AI developers with operational inputs on how to implement them in concrete scenarios. In addition, the lack of a clear ethical framework to refer to meant that this set of guidelines was not as robust and grounded as it might have been.[9]

More importantly, the ethical guidelines suggest a set of principles to be followed in AI development that seem to have little impact on companies. Some, such as Clearview AI[10] and Open AI,[11] have clearly violated not only ethical principles but also existing laws in order to bring their innovative products to market, reversing the model of responsible innovation.[12] It is therefore not surprising that the initial regulatory approach has shifted from ethical guidance to risk-based regulation.[13]

In a situation where the drive for innovation faces concerns about potential negative effects on a large scale, the logic of risk management seems to be the most appropriate way to strike a balance and define what risk is acceptable and to what extent. Moreover, it introduces an *ex ante* approach that makes it possible to prevent harmful applications from being placed on the market, rather than assessing *ex post* the correct behaviour of manufactures and deployers in complying with general principles.

The risk-based approach establishes procedural checks to be carried out, starting from the technology design phase and before the product is placed on the market. Although widely used in industrial regulation and mainly related to safety and security, this approach is also known and applied in the field of fundamental rights and, in a broader context, in relation to societal issues.[14] A clear example of this is data protection legislation, which starting from earlier generations[15] has followed a procedural approach focusing on the risks that each stage of data processing – from data collection to data erasure –

---

[4] See the decisions of the Garante per la protezione dei dati personali (the Italian data protection authority) on OpenAI's ChatGPT: Decision n. 112, 30th March 2023, doc. web n. 9870832 <https://www.garanteprivacy.it/web/guest/home/docweb/-/docweb-display/docweb/9870832> accessed 5 February 2024; Decision n. 114, 11th April 2023, doc. web n. 9874702 <https://www.garanteprivacy.it/web/guest/home/docweb/-/docweb-display/docweb/9874702> accessed 5 February 2024.

[5] See United States District Court, Southern District of New York, The New York Time Company v. Microsoft Corporation, OpenAI, Inc., OpenAI LP, OpenAI GP, LLC, OpenAI, LLC; OpenAI OPCO LLC, OpenAI Global LLC, OAI Corporation, LLC, and OpenAI Holdings, LLC <https://nytco-assets.nytimes.com/2023/12/NYT_Complaint_Dec2023.pdf> accessed 15 January 2024.

[6] See Charles D. Raab, 'Information Privacy, Impact Assessment, and the Place of Ethics' (2020) 37 Computer Law & Security Rev. 105404, https://doi.org/10.1016/j.clsr.2020.105404.

[7] The most common values listed in the adopted ethics guidelines can be classified within three main groups. The first consists of broad principles derived from ethical and sociological theory (common good, well-being, solidarity); the second group includes the principle of non-maleficence, the principle of beneficence, and the related broader notion of harm prevention (harm to social, cultural, and political environments); the last group includes openness, explicability and sustainability. For the analysis of these guidelines and their values, see Alessandro Mantelero, *Beyond Data: Human Rights, Ethical and Social Impact Assessment in AI* (T.M.C. Asser Press-Springer, 2022), https://doi.org/10.1007/978-94-6265-531-7 (open access), 97-101. See also Anna Jobin, Marcello Ienca, and Effy Vayena, 'The Global Landscape of AI Ethics Guidelines' (2019) 1 Nature Machine Intelligence 389-399; Thilo Hagendorff, 'The Ethics of AI Ethics: An Evaluation of Guidelines' (2020) 30 Minds and Machines 99-120.

[8] See also Linnet Taylor and Lina Dencik, 'Constructing Commercial Data Ethics' (2020) Technology and Regulation 1–10, https://doi.org/10.26116/techreg.2020.001.

[9] See also Mantelero, Beyond Data (fn.7), where the following four main shortcomings of this initial ethics-based approach were identified: uncertainty, heterogeneity, context dependence, and risks of a 'transplant' of ethical values.

[10] See Garante per la protezione dei dati personali (Italian GDPR Supervisory Authority), Ordinanza ingiunzione nei confronti di Clearview AI, 10th February 2022 <https://www.garanteprivacy.it/web/guest/home/docweb/-/docweb-display/docweb/9751362> accessed 12 December 2023; a summary of the decision in English is available here: https://www.garanteprivacy.it/home/docweb/-/docweb-display/docweb/9751323#english; CNIL (French GDPR Supervisory Authority). 2022. Restricted Committee Deliberation No. SAN-2022-019 of 17 October 2022 concerning CLEARVIEW AI <https://www.cnil.fr/sites/cnil/files/atoms/files/deliberation_of_the_restricted_committee_no_san-2022-019_of_17_october_2022_concerning_clearview_ai.pdf> accessed 12 December 2023.

[11] See fn. 4.

[12] See also Zoë Schiffer and Casey Newton, 'Microsoft Lays off Team That Taught Employees How to Make AI Tools Responsibly' (2023) The Verge, 14 March 2023 <https://www.theverge.com/2023/3/13/23638823/microsoft-ethics-society-team-responsible-ai-layoffs> accessed 10 January 2024.

[13] See Rec. 26, AI Act (Regulation (EU) 2024/1689 of the European Parliament and of the Council of 13 June 2024 laying down harmonised rules on artificial intelligence and amending Regulations (EC) No 300/2008, (EU) No 167/2013, (EU) No 168/2013, (EU) 2018/858, (EU) 2018/1139 and (EU) 2019/2144 and Directives 2014/90/EU, (EU) 2016/797 and (EU) 2020/1828 (Artificial Intelligence Act).

[14] See also Alina Wernick, 'Impact Assessment as a Legal Design Pattern—A "Timeless Way" of Managing Future Risks?' (2024) 3(2) Digital Society 29, paras 1.1. and 1.2. https://doi.org/10.1007/s44206-024-00111-4; Raab, 'Information Privacy, Impact Assessment, and the Place of Ethics' (fn. 6); European Union Agency for Fundamental Rights, 'Getting the future right – Artificial intelligence and fundamental rights' (2020) https://fra.europa.eu/en/publication/2020/artificial-intelligence-and-fundamental-rights. On the EU legislator's risk-based approach to the digital society and the broader, long-standing risk-based approach to industrial regulation, see Raphaël Gellert, 'The Role of the Risk-Based Approach in the General Data Protection Regulation and in the European Commission's Proposed Artificial Intelligence Act: Business as Usual?' (2021) 3(2) Journal of Ethics and Legal Technologies 15–33.See also Articles 2(2), 8(2), 43(3), 48, and Annex I, of the AI Act. For a broader perspective see Julia Black, 'Risk-Based Regulation: Choices, Practices and Lessons Being Learnt' in OECD (ed) In Risk and Regulatory Policy: Improving the Governance of Risk (OECD Publishing, 2010), 185–224; Maria Weimer, 'Risk as a Regulatory Idea' in Maria Weimer (ed) *Risk Regulation in the Internal Market: Lessons from Agricultural Biotechnology* (Oxford University Press, 2019), 19-46.

[15] See Viktor Mayer-Schönberger, 'Generational Development of Data Protection in Europe' in Philip E. Agre and Marc Rotenberg (eds) *Technology and Privacy: The New Landscape* (The MIT Press, 1997), 219–241.

may pose to individuals in terms of impact on privacy and fundamental rights.

As with ethics-based regulation, the risk-based approach can have many different nuances, including in relation to fundamental rights. These rights can be assigned varying degrees of importance in the risk-based approach, and their possible compromise accepted or not. More generally, risks can be weighed against the potential benefits of innovation, leading to various strategies in terms of the acceptance of risks.[16]

In the AI Act, the legislator opted for an 'acceptable' risk and primarily limited the scope of the Act to a closed list of cases that the law deems to pose a high risk. This pro-innovation view in framing the risk-based logic is consistent with a first-generation law such as the AI Act. As at the beginning of the industrial revolution and, later, at the dawn of the information society era, the initial regulatory approach balances risk mitigation with its cost in terms of a chilling effect on investment in innovation. In the early stages of innovation implementation, therefore, a kind of 'tolerance' by the legal system may be considered as an appropriate regulatory strategy, accepting a certain number of side effects for individuals and society in return for the future benefits that investment in new technologies can bring.

Viewed through the lens of technology and innovation regulation, the way the AI Act regulates AI industry is therefore not surprising. However, it departs partially from principles laid out in earlier debate on ethics and key guiding values, and pays limited attention to the protection of fundamental rights because the primary nature of the AI Act is product safety regulation.

Against this background, the Fundamental Rights Impact Assessment (FRIA) was a result achieved by the European Parliament vis-à-vis a Commission proposal that emphasised a human-centric approach and protection of fundamental rights but that failed to implement them adequately in the risk-based model outlined in the bill. As a result of this compromise, fundamental rights and the FRIA now play a role in various assessment procedures in the AI Act, not only in Article 27, where the FRIA is specifically regulated.

The FRIA is not an entirely new invention of the EU legislator in the AI Act. It is based on considerable experience of Human Rights Impact Assessment (HRIA), established at international level and implemented by companies in various contexts.[17] However, it should be noted that HRIA in AI is not traditional HRIA,[18] and this difference needs to be recognised in the implementation of the FRIA.

The design of models to identify and assess potential risks to fundamental rights remains the 'elephant in the room' of the AI Act. Lack of experience with HRIA in relation to AI and limited methodological debate suggest that a more in-depth analysis of possible tools for conducting FRIA in relation to AI applications is required.

To address the various issues and questions related to the FRIA, the following sections will analyse the legal roots and nature of the FRIA, the way the FRIA is framed in the AI Act, the methodological issues and key criteria in the design of the FRIA model template and its components, and the scope of application of the FRIA beyond the limits of the AI Act.

## 2. The roots and nature of the FRIA

In order to design the FRIA and its implementation appropriately, it is crucial to place it in the broader context of impact assessment methodologies and practices. Although this is not the place to discuss risk management theories, it is important to stress that the FRIA cannot be designed without drawing on the methodological criteria adopted in this field.[19] This is necessary not only in order to obtain scientifically accurate results, but also because the FRIA must be integrated with other risk assessment procedures[20] that AI operators must comply with, starting with the Conformity Assessment required by the AI Act but not limited to this piece of legislation.

With regard to the specific area of fundamental rights, the experience acquired in two key types of assessment need to be specifically taken into account when designing the FRIA: the Privacy Impact Assessment (PIA)/Data Protection Impact Assessment (DPIA) and the Human Rights Impact Assessment (HRIA). To avoid repeating considerations expressed elsewhere on the limitations of PIA/DPIA in fully capturing the impact of AI applications,[21] it is important to note that the DPIA has the same main features as the FRIA, as discussed below: (i) an *ex ante* approach, (ii) a rights-based focus on risk assessment, (iii) a circular iterative structure[22] that follows the product/service throughout its lifecycle,[23] (iv) an expert-based nature.[24]

---

[16] See, e.g., National Institute of Standards and Technology – NIST, 'Artificial Intelligence Risk Management Framework (AI RMF 1.0)', 4, https://doi.org/10.6028/NIST.AI.100-1 ("While risk management processes generally address negative impacts, this Framework offers approaches to minimize anticipated negative impacts of AI systems and identify opportunities to maximize positive impacts").

[17] HRIA is a process for identifying, understanding, assessing and addressing the adverse impacts of projects and activities on the enjoyment of human rights in relation to potentially affected rights-holders. See United Nations - Human Rights Council, 'Guiding Principles on Business and Human Rights: Implementing the United Nations 'Protect, Respect and Remedy' Framework', Resolution 17/4 of 16 June 2011. See also Nora Götzmann (ed) *Handbook on Human Rights Impact Assessment* (Edward Elgar Publishing, 2019); The World Bank, *Study on Human Rights Impact Assessments A Review of the Literature, Differences with other Forms of Assessments and Relevance for Development* (World Bank Group, 2013) <http://documents1.worldbank.org/curated/en/834611524474505865/pdf/125557-WP-PUBLIC-HRIA-Web.pdf> accessed 29 June 2024. For some examples of HRIA in relation to specific cases, see LKL International Consulting Inc., 'Human Rights Impact Assessment of the Bisha Mine in Eritrea' (2014) <https://media.business-humanrights.org/media/documents/files/documents/Nevsun_HRIA_Full_Report__April_2014_.pdf> accessed 27 June 2024; Kendyl Salcito, 'Kayelekera HRIA Monitoring Summary' (2015) <https://nomogaia.org/wp-content/uploads/2015/10/KAYELEKERA-HRIA-MONITORING-SUMMARY-10-5-2015-Final.pdf> accessed 20 February 2021.

[18] See Mantelero, Beyond Data (fn.7), 51-52, and 83-84.

[19] See also Vania Skoric et al., 'Roles of Standardised Criteria in Assessing Societal Impact of AI', 2024 IEEE Conference on Artificial Intelligence (CAI) (2024), https://doi.org/10.1109/CAI59869.2024.00220.

[20] In the field of law, and in particular in the EU legislation on the digital society, there is no distinction between the concepts of impact assessment and risk assessment. In this respect, the Framework Convention on Artificial Intelligence and Human Rights, Democracy and the Rule of Law adopted by the Council of Europe, available at <https://rm.coe.int/1680afae3c> accessed 4 July 2024, in Article 16 and in other provisions, also uses the hendiadys risk and (adverse) impact in relation to assessment. See also United Nations-General Assembly, 'Seizing the opportunities of safe, secure and trustworthy artificial intelligence systems for sustainable development' (2024), A/78/L.49, <https://documents.un.org/doc/undoc/ltd/n24/065/92/pdf/n2406592.pdf?token=B2t37k81jBeRWpleLp&fe=true> accessed 4 July 2024 ("Strengthening investment in developing and implementing effective safeguards, including risk and impact assessments, throughout the life cycle of artificial intelligence systems to protect the exercise of and mitigate against the potential impact on the full and effective enjoyment of human rights and fundamental freedoms"). Moreover, in other social sciences where the distinction exists, such as in the literature on social impact assessment, the many similarities between impact assessment and risk assessment suggest a merger, see Hossein Mahmoudi et al., 'A framework for combining social impact assessment and risk assessment' (2013) 43 Environmental Impact Assessment Review 1–8, https://doi.org/10.1016/j.eiar.2013.05.003. In line with the conclusion of other authors, see e.g. CLTC UC Berkeley Center for Long-Term Cybersecurity, 'Guidance for the Development of AI Risk and Impact Assessments' (University of California, Berkeley, 2021), 5 <https://cltc.berkeley.edu/wp-content/uploads/2021/08/AI_Risk_Impact_Assessments.pdf> accessed 4 July 2024, the terms risk assessment and impact assessment are used interchangeably in this article, with a preference for risk assessment when referring to general evaluation of risks.

The main difference between DPIA and FRIA is scope: FRIA does not focus on data protection, but considers all the potentially affected rights and freedoms.[25] Thus, it has the same scope as HRIA, with reference to fundamental rights.[26]

The argument made during the drafting of the AI Act that the FRIA was superfluous, given the presence of the DPIA and the wording of Article 35 of the GDPR, reveals some weaknesses taking into account the implementation of the DPIA. First, assessing impact on the various fundamental rights through the lens of the DPIA leads to the use of data protection categories to justify the final decision, which largely obscures the rationale behind the assessment in relation to these rights.[27] By contrast, the FRIA, like the HRIA, entails specific consideration of each relevant fundamental right, as defined in doctrine and case law, with more accurate and transparent results in terms of assessment. Second, looking at DPIA practice, it is evident that attention given to rights other than data protection is minimal and usually not well elaborated.[28]

In this context, it is also important to note that although HRIA is very close to FRIA in terms of focus, HRIA as part of business due diligence has mainly been used as an *ex post* response to critical situations.[29] This differs from the approach of the AI Act, where, as in the GDPR and the DSA, impact assessment is a mandatory obligation to be fulfilled before any innovative solution is implemented in the real world.

Another difference concerns the operational scope of HRIA, and FRIA in the AI Act. Traditional HRIA is mainly a policy tool that provides companies with an assessment of the potential impacts and a list of possible solutions to prevent or mitigate them,[30] leaving it up to the company to decide which solutions to adopt and to what extent to reduce these impacts. In contrast, under the AI Act, the FRIA is a mandatory assessment, the results of which must be used to prevent or mitigate risk.[31]

Finally, FRIA in AI differs from traditional HRIA in the nature of the situations assessed. While HRIA is usually applied in the context of industrial activities located in a specific territory and impacting on a wide range of human rights, including social rights, AI products are often globally distributed solutions that usually impact on a limited range of fundamental rights. Two partial exceptions to this distinction are smart cities projects, where AI systems are deployed in a specific territorial context, and LLMs, which can be used for a wide range of different purposes, potentially impacting on all fundamental rights.

Based on the considerations outlined here, FRIA cannot be developed simply by taking HRIA practices into account and copying HRIA models. The risk management obligations set out in the AI Act require a model that provides some degree of risk measurement in order to steer the design of AI applications towards less impactful solutions and to make AI operators accountable in terms of risk management. On the other hand, the more limited scope of many contextual AI applications reduces the complexity of traditional HRIA.[32]

Before analysing how the FRIA is framed in the AI Act and the associated requirements and obligations, it is also worth noting that the general approach to risk-based regulation adopted by the EU in the AI Act departs from the path set out in the GDPR. While the GDPR adopts a risk-based approach centred on fundamental rights and sets the threshold of high-risk as an insurmountable limit,[33] if the potential prejudice is not justified by prevailing interests according to the balancing test, the AI Act has a broader scope, not focused only on fundamental rights, and adopts as a general criterion the acceptability of risk in high-risk applications.[34]

Acceptability means that the risk may remain high, albeit mitigated,

---

[21] Mantelero, Beyond Data (fn.7), 20-25, and Alessandro Mantelero and Maria Samantha Esposito, 'An Evidence-Based Methodology for Human Rights Impact Assessment (HRIA) in the Development of AI Data-Intensive Systems' (2021) 41 Computer Law & Sec. Rev., https://doi.org/10.1016/j.clsr.2021.105561, Section 4. Data Protection Impact Assessment (DPIA), as established by Article 35 GDPR, is closely related to Privacy Impact Assessment (PIA), which exists in different experiences around the world and has been an important tool since the mid-1990s. On the origins of privacy impact assessment, see Roger Clarke, 'Privacy impact assessment: Its origins and development' (2009) 25(2) Computer Law & Security Review 123–129. Although, the existing differences between the concepts of privacy and data protection in different jurisdictions, in particular between common law and civil law systems, affect the scope of these assessments, which examine areas that do not fully overlap, from a methodological point of view there are broad similarities that have led to their being considered together for the purposes of this article. For a broader analysis, see also David Flaherty, 'Privacy impact assessments: an essential tool for data protection' (2000) 7(5) Priv. Law & Pol'y Rep. 45; David Wright, 'The state of the art in privacy impact assessment' (2012) 28(1) Computer Law & Security Review 54–61; David Wright and Paul De Hert (eds), Privacy Impact Assessment (Springer 2012); David Wright, Michael Friedewald, and Raphael Gellert, 'Developing and Testing a Surveillance Impact Assessment Methodology' (2015) 5(1) Int'l. Data Privacy Law 40–53.
[22] See also ISO, Risk management. Guidelines. ISO 31000 <https://www.iso.org/standard/65694.html> accessed 9 February 2024.
[23] See also fn. 20.
[24] For a more detailed analysis of these specific features see fn.18.
[25] Hereinafter the reference to fundamental rights is to be understood as a cumulative reference to both fundamental rights and freedoms.
[26] See also European Union Agency for Fundamental Rights <https://fra.europa.eu/en/about-fundamental-rights/frequently-asked-questions#difference-human-fundamental-rights> accessed 10 January 2023 ("The term 'fundamental rights' is used in European Union (EU) to express the concept of 'human rights' within a specific EU internal context. Traditionally, the term 'fundamental rights' is used in a constitutional setting whereas the term 'human rights' is used in international law. The two terms refer to similar substance as can be seen when comparing the content in the Charter of Fundamental Rights of the European Union with that of the European Convention on Human Rights and the European Social Charter.").
[27] See Mantelero and Esposito (fn. 21).
[28] Ibid.
[29] See, e.g., Kendyl Salcito and Mark Wielga, 'Kayelekera HRIA Monitoring Summary' (NomoGaia 2015) <https://nomogaia.org/wp-content/uploads/2015/10/KAYELEKERA-HRIA-MONITORING-SUMMARY-10-5-2015-Final.pdf> accessed 20 February 2021; LKL International Consulting Inc., 'Human Rights Impact Assessment of the Bisha Mine in Eritrea' (2014) <https://media.business-humanrights.org/media/documents/files/documents/Nevsun_HRIA_Full_Report__April_2014_.pdf> accessed 26 October 2023.
[30] See, e.g., Kuoni, 'Assessing Human Rights Impacts. India Project Report' (2014) <https://www.humanrights-in-tourism.net/publication/assessing-human-rights-impacts> accessed 15 January 2024.
[31] See also fn. 123.
[32] This latter consideration cannot be applied to smart cities and LLMs where more time-consuming analysis is required for the reasons outlined above.
[33] See Articles 35 and 36, GDPR and Article 29 Data Protection Working Party, 'Guidelines on Data Protection Impact Assessment (DPIA) and determining whether processing is "likely to result in a high risk" for the purposes of Regulation 2016/679' (2017) 18.
[34] See Article 9(5), AI Act. See also Article 35(1), DSA ("Providers of very large online platforms and of very large online search engines shall put in place reasonable, proportionate and effective mitigation measures, tailored to the specific systemic risks identified […]").

if justified by the potential benefits associated with AI.[35] This is in line with a view common to industrial risk-based regulation in sectors where hazards are inherent, such as the chemical industry. However, this general approach based on risk acceptability must be consistent with the protection of fundamental rights as enshrined in EU law.

For this reason, to the extent that the AI Act does not provide for specific exceptions with regard to fundamental rights, the potential high risks of prejudice to fundamental rights resulting from the impact of AI cannot be considered as acceptable.[36] This is without prejudice to the application of the traditional balancing test in the case of competing interests, which may lead to proportionate and necessary restrictions on fundamental rights.

The FRIA, like the DPIA in the GDPR, is therefore characterised by a rights-based approach[37] to risk assessment, as opposed to a pure risks/benefits approach, in which all the competing interests, including economic ones, are placed on the same level, with trade-off between them.[38]

Another important feature of the FRIA is its *ex ante* approach, which is common to other legal obligations on risk assessment in various fields and is also in line with the UN Guiding Principles on Business and Human Rights.[39] This not only prevents impacting applications from being available on the market, but also makes the FRIA a tool for AI design, in line with the by-design approach.[40] By assessing the potential risks associated with the different design options in product/service development, developers will be guided on how to better comply with fundamental rights and will discard solutions that are less protective.

However, like all risk assessments of situations that may evolve over time, the FRIA is not a one-shot prior evaluation, but has a circular iterative structure.[41] The traditional main phases of risk management (planning/scoping, risk analysis, risk prevention/mitigation) are repeated, as technological, societal and contextual changes impact on some of the relevant elements of a previous assessment.

Finally, FRIA is necessarily an expert-based assessment. Although guidelines and templates can facilitate assessment, the impact on fundamental rights cannot be fully automatised. Indeed, the relevant parameters to be considered are inevitably context-based and grounded on expert knowledge of fundamental rights (i.e. theoretical and case law developments).

This last element leads to reflection on the background of experts involved in the FRIA, their interaction with other advisory bodies (e.g. ethics committees), the various ways in which they may interact with the entities they have to advise (e.g., internal or external experts, the role of an internal AI supervisor,[42] the role of laypersons), and deliberation models (consensus-based approach or majority voting).

These issues have been analysed and addressed in detail elsewhere.[43] For the scope of this article, it is sufficient to refer to the conclusions given there: there is no one-size-fits-all solution, different contexts require different approaches depending on the nature of AI products/services, their impacts and the nature and structure of AI operators.

Regardless of the solution adopted, it is considered best practice to provide evidence of level of confidence in expert-based evaluations. This should demonstrate the reliability, relevance and up-to-dateness of the evidence used to support the assessment, the use of appropriate expertise in setting its variables, and the level of agreement between experts.[44]

## 3. The AI Act: the impact on fundamental rights in FRIA and Conformity Assessment

With the key elements of the FRIA outlined above, it is now important to define the relationship between this type of assessment and the broad Conformity Assessment required by the AI Act. In fact, impact on fundamental rights is addressed not only in the FRIA, under Article 27, but is also one of the elements of the Conformity Assessment, and must also to be carried out when GPAI[45] is used in the context of high-risk AI systems or general-purpose models that pose a systemic risk.[46]

---

[35] These potential benefits are emphasised in Rec. 4 AI Act.

[36] In this sense, Article 27 on the FRIA does not refers to acceptability. See also fn. 123.

[37] See also Google Spain SL and Google Inc. v Agencia Española de Protección de Datos (AEPD) and Mario Costeja González, Case C-131/12, 13 May 2014, ECLI:EU:C:2014:317; Scarlet Extended SA v Société belge des auteurs, compositeurs et éditeurs SCRL (SABAM), Case C-70/10, 24 November 2011, ECLI:EU:C:2011:771; Belgische Vereniging van Auteurs, Componisten en Uitgevers CVBA (SABAM) v Netlog NV, Case C-360/10, 16 February 2012, ECLI:EU:C:2012:85.

[38] It is also worth noting that the AI Act does not include cost-effectiveness among the criteria for FRIA and Conformity Assessment, mentioning it only in Article 50(2) with regard to the transparency obligations for providers of certain AI systems and GPAI models.

[39] See United Nations - Human Rights Council, Guiding Principles on Business and Human Rights (fn.17).

[40] The by-design approach, which is explicitly referred to in Article 25 of the GDPR in the context of data protection, has a long trajectory in the field of computer-human interaction and is not limited to data protection but also applies to other legal and ethical values. See Batya Friedman and Peter H. Kahn Jr., 'Human Values, Ethics, and Design', in Andrew Sears and Julie A. Jacko (eds) *The Human-Computer Interaction Handbook* (CRC Press, 2008), 1177-1201; Batya Friedman (ed), *Human Values and the Design of Computer Technology* (Cambridge University Press, 1997); Sarah Spiekermann, *Ethical IT Innovation: A Value-Based System Design Approach* (Taylor & Francis, 2016); Lee A. Bygrave, 'Hardwiring Privacy' in Roger Brownsword, Eloise Scotford and Karen Yeung (eds) *The Oxford Handbook of Law, Regulation, and Technology* (Oxford University Press, 2017) 754-775.

[41] This element is not particularly emphasised in the AI Act, but Article 27(2) states that "If, during the use of the high-risk AI system, the deployer considers that any of the elements listed in paragraph 1 has changed or is no longer up to date, the deployer shall take the necessary steps to update the information". Moreover, this view of impact assessment as a continuous iterative process is consistent with both general risk management theory and the general approach adopted in the AI Act, see Article 9(2) AI Act.

[42] See also Dutch Data Protection Authority - Department For The Coordination Of Algorithmic Oversight (DCA), 'AI & Algorithmic Risks Report Netherlands' (2024), 14 <https://media.licdn.com/dms/document/media/D4E1FAQGVGEABdGCzmA/feedshare-document-pdf-analyzed/0/1705644733016?e=1706745600&v=beta&t=F6xN7_WugEauBNmKTuwdJQNCOzvR2-Vv2wUr_hH6fnQ> accessed 20 January 2024.

[43] See Mantelero, Beyond Data (fn.7), Ch. 3.

[44] See Australian Institute for Disaster Resilience, 'National Emergency Risk Assessment Guidelines' (2020), para 6.7 <https://www.aidr.org.au/media/7600/aidr_handbookcollection_nerag_2020-02-05_v10.pdf> accessed 12 January 2024.

[45] On the notion of General-Purpose AI (GPAI) used in the AI Act, see Rec. 97, AI Act. According to Rec. 85, General purpose AI systems "may be used as high-risk AI systems by themselves or be components of other high-risk AI systems" and therefore the obligations established for high-risk systems, including impact assessment, are applicable to them. See also Rec. 97 ("Although AI models are essential components of AI systems, they do not constitute AI systems on their own. AI models require the addition of further components, such as for example a user interface, to become AI systems. AI models are typically integrated into and form part of AI systems.").

[46] See Article 55(1)(b), AI Act ("[providers of general-purpose AI models with systemic risk shall:] assess and mitigate possible systemic risks at Union level […]") and Art. 3 (65) (" 'systemic risk' means a risk that is specific to the high-impact capabilities of general-purpose AI models, having a significant impact on the Union market due to their reach, or due to actual or reasonably foreseeable negative effects on public health, safety, public security, fundamental rights, or the society as a whole, that can be propagated at scale across the value chain").

When comparing FRIA and Conformity Assessment, it is worth noting that they follow the same procedure based on risk identification, analysis, and prevention/mitigation,[47] as both are grounded on general risk management methodology.

Moreover, the focus on fundamental rights is present in both models, as part of a broader assessment in the Conformity Assessment, and as a specific objective of the FRIA. The FRIA identifies a specific type of fundamental rights impact assessment, as defined in Article 27, which is an obligation only for AI deployers, but the general tool of fundamental rights impact assessment, rooted in the HRIA, is also part of the Conformity Assessment, under Article 9, as an obligation for AI providers.[48]

In addition, the two assessments, that is the FRIA under Article 27 and the assessment of the impact on fundamental rights under the Conformity Assessment, are linked. The expertise of AI deployers can therefore reduce the burden for providers by contributing in the definition of risk management strategy.[49] This is based on the consideration that the level of expertise of deployers in the implementation of the FRIA enables them to address some residual risks to fundamental rights, which are therefore not addressed in the Conformity Assessment by the providers, as they are left to the deployers' capacities.[50]

On the other hand, the FRIA cannot be properly performed without information about the AI system from the AI provider. This information flow between provider and deployer is not properly framed in the AI Act. Article 27(2) states that the deployer (performing the FRIA) may "in similar cases, rely on previously conducted fundamental rights impact assessments or existing impact assessments carried out by provider" and several provisions – rather vague in their wording[51] – add elements that can be interpreted as including the disclosure of the impact on fundamental rights as carried out by the provider in the context of the Conformity Assessment. However, a clear and specific disclosure obligation would have been more effective in terms of integrated risk management.

Finally, according to Article 9(5), the outcome of the Conformity Assessment shall influence the design and development of the AI system to eliminate (more correctly prevent) or reduce relevant risks. Where risks cannot be prevented, appropriate mitigating measures and controls must be implemented, and risk information, in accordance with Article 13, and training must be provided to deployers. A similar range of measures can be considered in the case of FRIA, as risk management prioritises risk prevention, with mitigation considered only when prevention is not feasible.

Despite these similarities, there are some differences between the two forms of assessment. The most important is the standards-based approach adopted by the EU legislator to develop the Conformity Assessment.[52] Apart from criticisms related to standard setting in general,[53] specific concerns relate to the nature of the impact on fundamental rights and the competence of standardisation bodies in dealing with fundamental rights.

If we consider standards as predefined procedures to be applied in every case or to a class of cases, it is clear that the variability of the impact of AI on fundamental rights cannot be captured in assessment standards. The following three parameters at least allow for the potential impact on fundamental rights to be different and require a case-by-case evaluation: (i) the specific characteristics of the technology used; (ii) the context of use; (iii) the categories of persons potentially affected.[54]

However, standards can also be intended as methodological approaches that outline how to address specific issues rather than providing checklists or pre-defined detailed procedures, as it is the case with standards for risk management systems. Thus, it is possible to establish a standard for fundamental rights impact assessment by outlining the key phases of the assessment, setting methodological requirements for each of them, and defining a common approach to risk measurement, as is described in the following sections.

Against this background, standardisation bodies show a lack of expertise in the field of fundamental rights, as acknowledged by the European Commission.[55] In addition, conflicts may arise between Conformity Assessment standards and the FRIA methodology, as Article 27 allows deployers to define their own methodological approach, as is the case in the field of data protection with the DPIA.

Experience with the GDPR and Directive 95/46/CE shows it is possible to define best practices in impact assessment without necessarily establishing formal standards.[56] Similarly, in HRIA there are no given standards, but rather different best practices that enrich the tools used to tackle a variety of risk scenarios and impacts.

---

[47] In Article 9(2), the process is described as a three-stage model (risk identification, risk analysis consisting in risk evaluation and estimation, risk management). These different stages can be divided into sub-phases, and different assessment models group them in different ways, but always based on these three functions of identification, analysis, and management (risk prevention or mitigation). See also the G7's Hiroshima Process International Guiding Principles for All AI Actors, available at <https://digital-strategy.ec.europa.eu/en/library/hiroshima-process-international-guiding-principles-advanced-ai-system> accessed 28 June 2024, that explicitly refers in the first principle to "identify, evaluate, and mitigate risks across the AI lifecycle"; OECD, 'Advancing Accountability in AI' (2023) <https://www.oecd.org/en/publications/2023/02/advancing-accountability-in-ai_753bf8c8.html> accessed 28 June 2024, which uses a four-step model (scope definition, risk assessment, risk treatment, and risk management government), with more emphasis on the autonomy of post-assessment monitoring and management, within a principles-based assessment not limited to fundamental rights. See also fn. 123.
[48] See Article 43 and related provisions in Annex VI and Articles 17(1)(g) and 9(2).
[49] See also Rec. 93, AI Act.
[50] See Article 9(5) ("With a view to eliminating or reducing risks related to the use of the high-risk AI system, due consideration shall be given to the technical knowledge, experience, education, the training to be expected by the deployer").
[51] See Article 13(1) AI Act ("An appropriate type and degree of transparency shall be ensured with a view to achieving compliance with the relevant obligations of the provider and deployer set out in Section 3"). See also Articles 13 (2) on the instructions to be given to deployers ("High-risk AI systems shall be accompanied by instructions for use in an appropriate digital format or otherwise that include concise, complete, correct and clear information that is relevant, accessible and comprehensible to deployers") and 13(3)(b)(iii) on the content of those instructions on the characteristics, capabilities and limitations of performance of the high-risk AI system, including "any known or foreseeable circumstance, related to the use of the high-risk AI system in accordance with its intended purpose or under conditions of reasonably foreseeable misuse, which may lead to risks to the health and safety or fundamental rights referred to in Article 9(2)". See also Recs 65 and 72.

[52] See Article 40, AI Act.
[53] See also Michael Veale and Frederik Zuiderveen Borgesius, 'Demystifying the Draft EU Artificial Intelligence Act — Analysing the Good, the Bad, and the Unclear Elements of the Proposed Approach' (2021) 22(4) Computer Law Review International 97, 105, https://doi.org/10.9785/cri-2021-220402.
[54] An example of this is AI-powered video surveillance, where standardisation is difficult due to the inability to anticipate all possible scenarios. The use of continuous monitoring or motion-activated cameras, self-deletion of collected data, facial recognition, and data sharing with other entities, are just some of the possible elements that can change the impact of the use of this technology. Moreover, the rights at risk, the balancing of competing interests, and the resulting impact (including on various categories of vulnerable persons) may vary depending on the context, such as a neighbourhood with high street crime, the interior of a school with minors, or national borders. This does not preclude the possibility of developing specific standard metrics for assessing AI performance, see e.g. OpenAI et al., ' GPT-4 Technical Report' arXiv, 4 March 2024, https://doi.org/10.48550/arXiv.2303.08774 accessed 15 March 2024. These metrics and AI performance assessment can be useful to better identify and understand potential consequences in terms of impact on fundamental rights, but are not in themselves metrics for standardising the FRIA. They can be included in the preliminary data collection in the planning and scoping phase with regard to the inherent dimension of the AI system; see below Section 5.1.
[55] See European Commision, Draft standardisation request to the European Standardisation Organisations in support of safe and trustworthy artificial intelligence, 5 December 2022 <https://ec.europa.eu/docsroom/documents/52376?locale=en> accessed 25 February 2023.

Concerned about the difficulties faced by deployers and providers in dealing with risk assessment, the EU legislator and the EU Commission look favourably on the use of standards. Ad hoc standards, if set for Conformity Assessment including the impact on fundamental rights,[57] are likely to be replicated in the FRIA. This is even more likely given the narrow solution presented in Article 27(5), which mandates the AI Office to "develop a template for a questionnaire, including through an automated tool, to facilitate deployers in complying with their obligations under this Article in a simplified manner": questionnaires or templates to assist deployers can be useful, but without reducing the FRIA to a mere questionnaire-based exercise, which is contrary to its nature (see Section 5 below).

Given the positive experience of the DPIA in data protection and the role of the bodies empowered by the AI Act to provide guidance and templates, another way forward is possible. Conformity Assessment standards should exclude assessment of impact on fundamental rights, while providers and deployers could develop it, following methodological guidance provided by the European Artificial Intelligence Board,[58] the AI Office,[59] and national competent authorities,[60] but without formal standards. This will provide more flexibility and facilitate a contextual approach that will benefit AI providers and deployers in properly addressing their assessment obligations.

Whatever strategy prevails, it will be necessary to develop a more robust methodological approach to assessing the impact of AI on fundamental rights. This study aims to contribute to this effort. In this respect, although a common methodology for this assessment in the FRIA and in the Conformity Assessment can be designed, its development under Article 9 in the context of Conformity Assessment may suffer from certain limitations.

The first major difference with the traditional approach adopted in HRIA/FRIA is the focus on the AI product alone. This overlooks the fact that these products operate in a context, and that this contextual dimension, especially with regard to impact on individuals, is crucial for risk prevention.[61] Article 9(3), limiting the risks to those which "may be reasonably mitigated or eliminated through the development or design of the high-risk AI system, or the provision of adequate technical information", does not recognise that AI systems are often socio-technical systems. The design to consider is therefore not only the design of the AI system, but also[62] the design resulting from the interaction and mutual modification that these systems generate in society.

Thus it is not only the "development or design of the high-risk AI system" that is relevant but also the conditions of the context of use, which in some cases can also be appropriately modified to prevent risks, without changing the AI design. For example, there is a difference between the use of an AI decision support system by competent public authorities in the context of humanitarian emergencies and the use of the same system under normal conditions. The state of stress of all the people involved in the first scenario can exacerbate data quality, poor human-AI interaction, and biases.

Another example, which does not depend on the human contextual factor but on the technological factor, concerns access to health services based on screening programmes. In this case, although the AI system works correctly, its interaction with poor diagnostic instruments exacerbates false positives and negatives, again demonstrating that the risks associated with AI need to be assessed in context and not just in terms of the design of the system.

Given these considerations, it can be concluded that the three-layered risk assessment structure adopted by the AI Act, that is an assessment close to technological assessment (Article 7),[63] Conformity Assessment (Article 9), and FRIA (Article 27), requires methodological reflection in order to harmonise the risk management approach with regard to fundamental rights.

## 4. The FRIA as set out in Article 27 of the AI Act

The FRIA was introduced by the European Parliament during the legislative process of the AI Act as an obligation for AI deployers, a category added by the Parliament to bridge the gap between AI providers and end users, emphasising the role that deployers can actively play in the contextual use and customisations of AI systems.[64] In line with general risk theory, the burden of risk management is therefore shared proportionally between AI providers and deployers, according to the actual risk introduced into society and their respective power to manage it.[65]

The deployer's role relates to the relevant contextual component of risk management in the specific use of an AI system. Some risk elements, such as specific vulnerabilities of the individuals concerned, cannot be foreseen or properly managed by the provider at a general level. Against this background, the condition for risk management by the deployer is feasibility: the system should be sufficiently accessible and 'customisable' by the deployer and adequate risk information should have been made available by the provider.

---

[56] This approach also avoids the limitations of a top-down exercise, which is often not particularly inclusive. See also the notion of guardrails used by Urs Gasser and Viktor Mayer-Schönberger, Guardrails: Guiding Human Decisions in the Age of AI (Princeton University Press, 2024), 104, 186-187, and 189

[57] See Rec. 121, AI Act ("The common specification should be an exceptional fall back solution to facilitate the provider's obligation to comply with the requirements of this Regulation […] when the relevant harmonized standards insufficiently address fundamental rights concerns").

[58] See Art. 66, b, e and l, AI Act.

[59] See Art. 27.5, AI Act.

[60] See Art. 70.8, AI Act.

[61] See in this sense the explicit reference to the context of use in risk assessment in Council of Europe, AI Framework Convention (fn. 20), Article 16(2)(a) ("[measures for the identification, assessment, prevention and mitigation of risks] shall be graduated and differentiated, as appropriate and: a. take due account of the context and intended use of artificial intelligence systems […]").

[62] See also Art. 9.5.a, AI Act.

[63] See also Rec. 48 AI Act. See Rasmus Øjvind Nielsen et al., 'Ethical Assessment of Research and Innovation: A Comparative Analysis of Practices and Institutions in the EU and selected other countries. Deliverable 1.1' (2015) <https://satoriproject.eu/media/D1.1_Ethical-assessment-of-RI_a-comparative-analysis.pdf> accessed 12 December 2023 ("technology assessment (TA) is a form of impact assessment that is specifically developed to assess impacts of a new technology. TA investigates the potential and actual effects of new technologies on industry, the environment and society, evaluates such effects and develops instruments to steer technology development in more desirable directions. TA makes such assessments on the basis of known or potential applications of the technology. It pays special attention to consequences that are unintended, indirect or delayed."). See also Armir Grunwald, 'The Objects of Technology Assessment. Hermeneutic Extension of Consequentialist Reasoning' (2020) 7(1) Journal of Responsible Innovation 96–112. https://doi.org/10.1080/23299460.2019.1647086; Armin Grunwald, Technology Assessment in Practice and Theory (Routledge, 2018); Armin Grunwald, 'Technology Assessment: Concepts and Methods' in Anthonie W.M. Meijers (ed) Philosophy of Technology and Engineering Sciences. Handbook of the Philosophy of Science, vol. 9 (Elsevier, 2009), 1103-1146. Technology Assessment therefore differs from Conformity Assessment and FRIA in that it focuses on the technological advance and its future implications in relation to a specific technological area, whereas the others focus on a specific, existing application of technology in that area.

[64] See also Rec. 93, AI Act.

[65] There are risks both in the development of AI and in its use in a concrete scenario. The provider is in the best position to manage the former and can foresee the latter, as well as the reasonable possible AI applications. On the other hand, the deployer can contribute to the management of the latter. See also Article 60(4)(h) of the AI Act on testing of high-risk AI systems in real world.

While the European Parliament's proposal included many key elements of FRIA, it did not outline the key parameters for assessment[66] and did not emphasise the by-design approach that is common in technology regulation. However, the Parliament's proposal provided for a higher level of detail on FRIA compared to the final text of the AI Act, where some elements are implicit, due to a mistaken attempt to reduce the burden on deployers. These elements are the mitigation plan,[67] consideration of vulnerability,[68] and a clear description of the components of this assessment.

As discussed below, the lack of explicit reference to these and other key elements does not result in a kind of simplified version of the FRIA, as this would be contrary to the inherent nature of this type of assessment, would not provide the adequate level of protection of the fundamental rights enshrined in the EU Charter of Fundamental Rights, and would diverge significantly from best international practice in this field.[69] In this respect, the FRIA can be effective only if carried out as it should be in its complete form, and the decision not to include details of important components of the FRIA in the final text only has the effect of complicating compliance with the AI Act by deployers.

Another important part of the European Parliament's proposal concerned the role of participation in impact assessment.[70] Unfortunately, although EU institutions emphasise participation in many contexts, it seems difficult to translate this attitude into legal obligations when it comes to the digital society. As in the GDPR, the AI Act does not give due attention to participation in assessment procedures, contrary to best practices in impact assessment.[71]

However, the main difference between the European Parliament's proposal and the adopted text concerns the scope of the FRIA. Under pressure from the other two co-legislators, it was restricted to a limited area, whereas the text proposed by the Parliament referred to all high-risk AI systems as defined in Article 6(2), with the sole exception of systems used for management and operation of critical infrastructure.[72]

The final text maintains this exception but significantly narrows the general scope of the FRIA, which now only covers (i) deployers that are "bodies governed by public law" and "private entities providing public services",[73] and (ii) AI systems used to evaluate the creditworthiness of natural persons or for credit scoring (with the exception of AI systems used for the detection of financial fraud), and for risk assessment and pricing in life and health insurance.[74]

Although this narrow scope of the FRIA is less satisfactory from the perspective of the protection of fundamental rights and creates an imbalance between the general obligation of providers to assess the impact on fundamental rights of all high-risk AI systems under the Conformity Assessment procedure and the specific obligation of deployers, it does not prevent the adoption of a broader use of this instrument based on the obligation to protect fundamental rights established at EU and national level, and facilitating the accountability of AI operators in this respect.[75]

Looking at the adopted text, the first paragraph of Article 27 places the FRIA, as defined by the AI Act, in the general HRIA/FRIA tradition, but in the form of a prior assessment, and no longer as a mere policy tool to respond to criticisms raised.

Given the nature of fundamental rights and the level of protection afforded to them by the Charter of Fundamental Rights of the European Union and national constitutional charters, this assessment must necessarily avoid prejudice to them. This means that the FRIA cannot merely be a final check with no influence on AI design. On the contrary, potential impacts must be properly addressed in order to meet the obligations to protect fundamental rights.

Due to the link between potential risk and system design,[76] it is recommended that this assessment be performed from the early stages of definition of the deployer's strategy for the use of a given AI system. It should also be repeated whenever changes are made to the system's deployment.[77]

Another important element in framing the FRIA, set in the first paragraph, is its objective, which is to assess "the impact on fundamental rights that the use of such system may produce". The focus is on the impact. This is a broad notion encompassing any kind of restriction or prejudice to fundamental rights.

Similar conclusions can be drawn from Article 27(1)(f), which refers

---

[66] The Brazilian legislative debate on the regulation of AI, for example, has taken a different approach by attempting to outline key criteria for impact assessment. See Senado Federal, Projeto de Lei n° 2338, de 2023, Dispõe sobre o uso da Inteligência Artificial. See also, Senado Federal, Emenda -CTIA (Substitutivo) ao Projeto de Lei 2.338/2023. See also Hana Mesquita et al., 'Regulating Artificial Intelligence in Brazil: The Contributions of Critical Social Theory to Rethink Principles' (2024) Technology and Regulation 2024 73–83. https://doi.org/10.26116/techreg.2024.008.

[67] See Article 29a(1)(h), AI Act EP (Proposal for a Regulation of the European Parliament and of the Council laying down harmonized rules on Artificial Intelligence (Artificial Intelligence Act) and amending certain union legislative acts 2021/0106(COD) DRAFT, Final draft as updated on 21/01, EP Mandate) which refers to "a detailed plan as to how the harms and the negative impact on fundamental rights identified will be mitigated".

[68] See Article 29a(1)(f), AI Act EP, referring to "specific risks of harm likely to impact marginalised persons or vulnerable groups". Although the definition of affected categories can be difficult and requires a contextual analysis, vulnerability is a component of any impact assessment focused on rights.

[69] See, e.g., United Nations - Human Rights Council. 2011. Guiding Principles on Business and Human Rights (fn. 17) and United Nations - Human Rights Council, 'The Right to Privacy in the Digital Age: Report of the United Nations High Commissioner for Human Rights' (15 September 2021) UN Doc A/HRC/48/31, <https://documents.un.org/doc/undoc/gen/g21/249/21/pdf/g2124921.pdf?token=LuFsNwgP4wDqSsTaBR&fe=true> accessed 12 February 2024.

[70] See Article 29a(4), AI Act EP ("In the course of the impact assessment, the deployer, with the exception of SMEs, shall notify national supervisory authority and relevant stakeholders and shall, to best extent possible, involve representatives of the persons or groups of persons that are likely to be affected by the high-risk AI system, as identified in paragraph 1, including but not limited to: equality bodies, consumer protection agencies, social partners and data protection agencies, with a view to receiving input into the impact assessment. The deployer shall allow a period of six weeks for bodies to respond. SMEs may voluntarily apply the provisions laid down in this paragraph. In the case referred to in Article 47(1), public authorities may be exempted from this obligations."). It is worth noting that the proposed provision already suggested a rather limited form of participation, closer to a consultation, and without defining the level of transparency and information provided to stakeholders.

[71] See below Section 5.3. The Parliament also proposed a transparency obligation for the AI deployer in certain cases, see Annex VIII, Section B, AI Act EP ("5. a summary of the findings of the fundamental rights impact assessment conducted in accordance with Article 29a").

[72] See Article 29(a)(1) and Annex III, area 2, EP AI Act.

[73] See also Rec. 96, AI Act ("Services important for individuals that are of public nature may also be provided by private entities. Private operators providing such services of public nature are linked to tasks in the public interest such as in the area of education, healthcare, social services, housing, administration of justice").

[74] See Annex III, 5(b) and (c), AI Act.

[75] See below Section 6.

[76] See also, for a systemic interpretation, Article 9(3) which refers to the risk that can be "mitigated or eliminated through the development or design of the high-risk AI system".

[77] See Rec. 96, AI Act ("The impact assessment should be performed prior to deploying the high-risk AI system, and should be updated when the deployer considers that any of the relevant factors have changed"). This is also supported by Article 27(4), which combines the FRIA and the DPIA in their common areas, with the DPIA being carried out from the early stages of data processing design.

to the materialization of a risk and the measures to be taken in the event of this happening. This provision must be interpreted in the light of the wording of this article and its consistency with risk management theory, as follows.

First, the materialization of these risks referred to in Article 27(1)(f) is distinct from the materialisation of harm[78] and *ex post* remedies. It therefore lies in the realm of risk assessment and management.[79] When a deployer designs a specific use of an AI system, the risk is introduced into society and materialises at that moment,[80] giving rise to the obligation to take appropriate measures to address it. The emergence of specific elements in the design of AI deployment that may have a negative impact on fundamental rights therefore constitutes the materialisation of the risk.[81] Using an *a contrario* argument, it would not make sense in the risk management-oriented AI Act to impose an obligation to define "the measures to be taken in the case of materialisation of those risks", considering materialisation as the actual occurrence of harm, but not taking appropriate measures when the risk is foreseen before it causes actual harm. Contrary to the preventive nature of the risk management approach, the response to risk would only be in the form of remedies.

Second, the FRIA is qualified by the AI Act as an instrument to be adopted "prior to deploying a high-risk AI system". This excludes an *ex post* approach. In the same vein, the reference in Article 27(1) to the assessment of the impact that the use of an AI system "may produce" and the likelihood criterion ("categories of natural persons and groups likely to be affected" and "risks of harm likely to have an impact") highlight the preventive and predictive nature of the analysis to be carried out.[82]

Article 27's purpose, as revealed during legislative debate, and systemic interpretation, the AI Act's risk-based approach focusing on an *ex ante* instrument, and the general theory of FRIA/HRIA, lead to the conclusion that the FRIA is a prognostic assessment and management exercise, similar to the DPIA.[83]

In line with HRIA and risk management in general,[84] the FRIA must include at least (i) a planning and scoping phase, focusing on the main characteristics of the product/service and the context in which it will be placed;[85] (ii) a data collection and risk analysis phase, identifying potential risks and estimating their potential impact on fundamental rights;[86] (iii) a risk management phase, adopting appropriate measures to prevent or mitigate these risks and testing their effectiveness.[87]

These various components are examined in the following section. But it is worth noting that the FRIA cannot be reduced to a mere descriptive exercise, in which potentially affected rights are outlined in general terms and some measures are proposed, without any evidence of the link between these two in terms of the adequacy and effectiveness of the measures in reducing the estimated levels of risk.

The questionnaire to be developed by the AI Office under Article 27 (5) may therefore assist deployers in fulfilling certain obligations of the FRIA, particularly in the planning and scoping phases, as well as some aspects of data collection during the assessment phase. However, a purely questionnaire-based approach, and even worse its automation (which requires a high degree of uniformity), cannot fully capture the contextual nature of the FRIA and needs to be integrated with a methodology for risk quantification and management.[88]

As regards the life cycle of AI systems, the AI Act adopts for FRIA the circular iterative approach common in risk assessment,[89] where the measures adopted to manage risks need to be revised according to the technological and contextual changes. This is clear in the procedural aspects described in Article 27(2), although the reference to updating information can be misleading on a quick reading,[90] as the update

---

[78] On the notion of harm in Article 27, see also fn. 125.

[79] In this respect, both internal governance and the complaint mechanism – to which Article 27(1)(f) refers as examples of possible measures – can be used as relevant tools at the design stage. Here, complaints may relate to previous similar uses of AI on which the new system builds, or be part of a participatory approach in which the public administration presents an AI-based solution and this raises specific complaints about potential risks. See also fns. 73 and 82.

[80] To give a paradigmatic example, the risk associated with the use of the atomic bomb materialised during the experiments for its realisation, without the need for its concrete use; when it was actually used, this caused the materialisation of the harm. For a similar, more recent and AI-related example, see research into the development of robots with high mobility skills (e.g., Boston Dynamics's robots), which has raised concerns about the risks associated with their possible future use in military or other critical situations with potential harm to humans. From the point of view of systemic interpretation, it is worth noting as the term 'materialisation' (of harm) was used in Article 7(2), point (c) in both the Commission and Council proposals and was replaced by the European Parliament and in the final text by 'likelihood' of harm. This confirms the probabilistic and prior assessment nature of the interpretation to be applied to this concept. Moreover, the concept of materialisation refers to something that becomes real, and a risk becomes real when the nature of the risk, the interests impacted, the people affected, and the potential consequences can be foreseen; see Merriam Webster, online dictionary, Materialize ("to come into existence") <https://www.merriam-webster.com/dictionary/materialize> accessed 9 April 2024, and Cambridge Dictionary ("to become real or true") <https://dictionary.cambridge.org/dictionary/english/materialize> accessed 9 April 2024.

[81] For example, if a public administration plans to deploy an AI-powered chatbot to provide online information about administrative procedures and rights in the field of family law, the associated risk materialises and needs to be properly mitigated. The deployer is introducing new forms of interaction with end-users entailing specific risks in terms of potentially incorrect or incomplete information or incorrect human-machine interaction. See, for similar applications, Aroged, 'ChatGPT: Ministry of Justice Will Use GPJ to Respond to Citizens in Portugal' (*Aroged*, 17 February 2023) <https://www.aroged.com/2023/02/17/chatgpt-ministry-of-justice-will-use-gpj-to-respond-to-citizens-in-portugal/> accessed 7 February 2024; the application is available at <https://justica.gov.pt/en-gb/Servicos/Justice-Practical-Guide-Beta-Version> accessed 7 February 2024.

[82] In some circumstances, residual risks cannot be excluded; therefore, the assessment should also consider these risks and related complementary *ex post* measures to be adopted (e.g. compensation). The internal governance and complaint mechanisms referred to in Article 27(1)(f) can also be used in this case.

[83] Their common nature and ex ante procedural approach is recognised in Article 27(4). According to this provision, the FRIA "shall complement that data protection impact assessment", if some of the obligations under the FRIA are already fulfilled by the DPIA.

[84] See, for example, ISO, Risk management. Guidelines. ISO 31000 (fn. 22), which identifies the following three main phases, combined with three complementary tasks (recording & reporting; monitoring & review; communication & consultation): (i) scope, context and criteria; (ii) risk assessment (risk identification, risk analysis, risk evaluation); (iii) risk treatment.

[85] See Article 27(1), letters a), b), and c).

[86] See Article 27(1), letters c) and d). The identification of "the categories of natural persons and groups likely to be affected" is relevant both as a contextual element (who is impacted?) and as an element relating to the rights impacted (which rightsholders are affected?).

[87] See Article 27(1), letters e) and f).

[88] See Section 5.

[89] See Rec. 96, AI Act.

[90] The information that has changed relates to the factors that shape the FRIA and its outcome. The obligation set out in Article 27, paragraph 1, and the last part of paragraph 2, therefore necessarily lead to a revision of the FRIA in the case of changes relevant to the assessment, in line with impact assessment practice.

concerns the impact assessment, as confirmed by Recital 96.[91]

With regard to the availability of the assessment results, as mentioned above, the EU legislator departs from the traditional approach to impact assessment based on transparency. Some relief is provided by the obligation to notify the results of the assessment to the market surveillance authority;[92] this may increase the level of commitment by deployers compared to the experience with the DPIA in the GDPR, where any control over the impact assessment is left to the initiative of the Supervisory Authorities or arises in the event of legal action.

Finally, regarding the level of enforcement of FRIA obligations provided for by the AI Act, the Act does not introduce specific administrative fines, leaving it to the Member States to establish them in accordance with Article 99.[93] This may pose a significant risk of lack of effective protection of fundamental rights, such as has happened in some countries with regard to the exemption of the application of GDPR sanctions to the public sector.[94] Unfortunately such a restriction could also be based on Article 99(8) of the AI Act[95] and would undermine the impact of the FRIA, which includes the bodies governed by public law as one of its main targets.

However, given the importance of fundamental rights, the penalties for providers of high-risk AI systems that do not comply with their obligations (Articles 99(4)(a) and 16), which include an assessment of the impact on fundamental rights, and the distribution of risk management obligations between AI providers and deployers provided for in Article 27, it would be inconsistent with the purpose of the AI Act and the level of protection granted to fundamental rights by the EU legal system if Article 27 were not adequately accompanied by effective, proportionate and dissuasive sanctions.[96]

## 5. From obligations to methodology: key elements for a model template

Although fundamental rights impact assessment is not a new approach to risk management in rights protection, its implementation in the specific field of AI, with the peculiarities involved, is a recent development.[97] A first contribution to this was made by The Danish Institute for Human Rights in 2020.[98] In addition, the broad echo of the debate on the ethics of AI has also led a number of actors to develop core principles to be implemented in their AI systems in various ways, mainly through awareness-raising questionnaires.

It is only in recent years that more attention has been paid to risk-based methodologies, partially as a result of the shift from a purely ethical approach to AI regulation. The latter has its roots in product safety regulation, centred on a risk-based approach, risk assessment and risk management methodologies. However, this approach is not new in the human rights field, where human rights due diligence based on HRIA has been at the core of the UN Guiding Principles on Business and Human Rights for many years.[99]

This distinction between awareness-rising models and risk-based models for fundamental rights impact assessment, where the latter are fewer and still at developmental stage, refers to the main characteristics of the models considered, since a combination of elements from these different models is often present. Thus, questionnaires and an aim to raise awareness of the possible consequences of AI are common to all models, as is segmentation of the AI process to take account of the specificities of the different stages in terms of risk management. Finally, elements of risk quantification are also included in the models that do not properly develop a risk assessment centred on levels of risk.

All these models take a procedural approach because of the need to identify, assess and mitigate potential risks. However, when defining the FRIA methodology,[100] the importance of the contextual dimension must be considered. In this respect, the key question concerns the need to use an AI-based system rather than alternative possible solutions.

An example of this is provided by the Algorithmic Impact Assessment model developed by the Government of Canada,[101] which in its section on Reasons for Automation asks whether alternative non-automated processes were considered[102] with the following two sub-questions: If non-automated processes were considered, why was automation identified as the preferred option? What would be the consequence of not using the system?[103]

In terms of the procedural approach to FRIA, asking this key question at the beginning of the model, immediately after the general description of the AI system, is crucial to assessing the need for the use of AI in light

---

[91] This conclusion in favour of a coherence and continuous risk management is also confirmed by the last part of paragraph 2, which considers previous assessments (either carried out by the deployer or the provider) as a basis for further analysis, emphasising the continuity of the impact assessment cycle.

[92] See Article 27(3), AI Act.

[93] See Article 99(1), AI Act ("Member States shall lay down the rules on penalties and other enforcement measures, which may also include warnings and non-monetary measures, applicable to infringements of this Regulation by operators"). See also Rec. 168 ("Member States should take all necessary measures to ensure that the provisions of this Regulation are implemented, including by laying down effective, proportionate and dissuasive penalties for their infringement […] In order to strengthen and harmonise administrative penalties for infringement of this Regulation, the upper limits for setting the administrative fines for certain specific infringements should be laid down.").

[94] See, e.g., Article 77 of the Spanish data protection law (Ley Orgánica 3/2018, de 5 de diciembre, de Protección de Datos Personales y garantía de los derechos digitales).

[95] See Article 99(8), AI Act ("Each Member State shall lay down rules on to what extent administrative fines may be imposed on public authorities and bodies established in that Member State").

[96] See Article 99(1), AI Act ("the penalties provided for shall be effective, proportionate and dissuasive"). See also, e.g., ECJ, C-81/12, Asociaţia Accept v Consiliul Naţional pentru Combaterea Discriminării, 25 April 2013, paras 54-73, ECLI:EU:C:2013:275; ECJ, Lindqvist, C-101/01, 6 November 2003, paras 87-88, ECLI:EU:C:2003:596. In the worst-case scenario, in the absence of adequate sanctions for non-compliance with Article 27, the role of the DPIA could be reconsidered by going beyond the prevailing restrictive practice and fully implementing its fundamental rights impact assessment component; see below Section 6.

[97] This section builds on a broader study conducted in 2022 on the methodological approach to risk assessment in AI with respect to the potential impact of AI on human rights; see Mantelero, Beyond Data (fn.7), Ch 2.

[98] See The Danish Institute for Human Rights, 'Guidance on HRIA of Digital Activities' (2020) <https://www.humanrights.dk/publications/human-rights-impact-assessment-digital-activities> accessed 18 June 2021. Although not specifically focused on AI, this guidance opened the debate on the use of HRIA outside its traditional domain and in relation to digital technologies.

[99] See also Nora Götzmann, 'Introduction to the Handbook on Human Rights Impact Assessment: Principles, Methods and Approaches', in Götzmann, Handbook on Human Rights Impact Assessment (fn. 17), 2–30.

[100] We refer to FRIA here as the Fundamental Rights Impact Assessment under the AI Act, including both the assessment under Article 27 and the fundamental rights component of the Conformity Assessment, as they both have to address the same methodological issues, as far as fundamental rights are concerned.

[101] The model is available at <https://open.canada.ca/aia-eia-js/?lang=en>, and the methodological note at <https://www.canada.ca/en/government/system/digital-government/digital-government-innovations/responsible-use-ai/algorithmic-impact-assessment.html> accessed 11 February 2024.

[102] See also UNESCO, 'Ethical Impact Assessment. A Tool of the Recommendation on the Ethics of Artificial Intelligence' (2023) <https://unesdoc.unesco.org/ark:/48223/pf0000386276> accessed 10 October 2023, question 2.2.1 ("Has careful consideration been given to non-algorithmic options which may be used to achieve the same goal? If so, why is the option involving an AI system favoured?").

[103] See also the following section.

of its potential benefits and risks.[104] The same question should be repeated at the end of the impact assessment in order to evaluate whether, on the basis of the potential impacts envisaged, alternative solutions that do not rely on AI entail a lower risk while achieving the same results.

As mentioned above, the three main blocks of the FRIA methodology to be reflected in a model template are: (i) planning and scoping, and risk identification, (ii) risk analysis, and (iii) risk management.[105] The different types of models are considered in the following subsections and the contribution they can make to the design of each of these blocks of FRIA methodology is examined. While the procedural approach is common to the whole FRIA, the awareness-raising models can contribute more to understanding how to design the planning and scoping and risk identification phase of FRIA, and the models more centred on risk analysis and quantification can be used to develop the risk analysis and management phase.

### 5.1. First FRIA block: awareness raising and risk identification

In line with the structure of Article 27, the first objective of the assessment should focus on contextualising the AI system and its use, and identifying potentially affected categories (rightsholders).[106] Given the place of FRIA obligations in the broader context of existing EU law and fundamental rights principles, this planning and scoping phase must first address the legal acceptability of the AI option.

Two main areas need to be explored at this stage: the inherent dimension of the AI system and the contextual dimension. Both contribute to analysing the problem, considering alternative solutions, and, if AI is the best option, defining the goal of the AI system.

With regard to the inherent dimension of the AI system, awareness-raising aims to investigate (i) the specific needs to be addressed; (ii) the relevance of adopting an AI-based approach; (iii) the role of the AI-based solution in addressing the identified needs. It will therefore also cover the description of the nature of the AI systems, including data flows and data processing purposes.

In this respect, a possible model for this phase has been developed by the Dutch Ministry of the Interior and Kingdom Relations. It uses a questionnaire-based approach, listing many different questions at this awareness-raising phase[107] and combining them with a scenario-based analysis, where the questions are addressed in the light of possible alternative scenarios.

It is worth noting that this Dutch model includes some requirements in the planning and scoping phase, such as the existence of a legal basis that "explicitly and clearly" allows for the use of an algorithm and "renders this use sufficiently foreseeable",[108] that are neither required by the AI Act nor necessary from a general perspective of fundamental rights protection.[109] There is also a reference to the values driving[110] the decision to use AI. This is important, but operates at a different level of societal impact.[111]

In terms of methodology, the Dutch model adopts a Why-What-How approach, which can be of help in the decision-making process (why do we need AI? What AI? How will it work?), but does not fully fit with the logic of FRIA. The Why-What-How approach primarily concerns decisions related to needs and product features, but impact on fundamental rights runs through all three stages and even covers elements distributed over all three. In this respect, the most common risk management design is not linear through the Why-What-How stages, but circular, centred on planning/scoping, analysis and management.[112]

Another weakness of this model in defining the inherent dimension of the AI solution concerns the consideration of several design elements in their general dimension, such as product features,[113] biases and data quality,[114] and data security.[115] While the focus on these aspects is relevant in terms of raising awareness, it lacks a specific link to potentially affected rights in the model proposed, as should be the case in a FRIA model.[116]

With regard to the contextual dimension of AI solutions to be explored in the awareness raising stage, given the nature of FRIA, it is not limited to the identification of potentially affected rights and rightsholders (without quantifying the impact, which is the objective of the following phase), but it also includes a preliminary analysis of the relevant elements of the specific fundamental rights systems. These need to be properly contextualised according to the way in which they are shaped by legal provisions and case law, including with different nuances at national level. This also comprises the obligations and legal requirements already in place that mitigate or prevent potential risks (for example, the GDPR's provisions on data quality in relation to the data used to fine-tune AI).

In addition, the context of use is relevant in terms of the interaction between the AI system and the socio-technical environment in which it operates. This interaction relates to both the characteristics of the system and the characteristics of the surrounding environment, where the latter may be more difficult to control and modify through risk management.

To deal with the concrete design and contextual use of an AI system, and to identify potentially impacted rights, various approaches are possible and have been established in technology risk assessment over the decades.[117] These include brainstorming analysis, knowledge-based approaches based on past cases, failure models, checklists and various other methods. However, in the field of AI-related risk assessment, these methods are relatively unexplored and the models proposed tend to rely on checklists.[118]

---

[104] See also Council of Europe. 2019. Guidelines on artificial intelligence and data protection adopted by the Committee of the Convention for the Protection of Individuals with regards to Processing of Personal Data (Convention 108) on 25 January, para 2.9 <https://rm.coe.int/2018-lignes-directrices-sur-l-intelligence-artificielle-et-la-protecti/168098e1b7> accessed 10 January 2024, ("In order to enhance users' trust, AI developers, manufacturers and service providers are encouraged to design their products and services in a manner that safeguards users' freedom of choice over the use of AI, by providing feasible alternatives to AI applications.").
[105] See the previous section.
[106] See Article 27(1)(a), (b) and (c), AI Act. See also above fn. 54.
[107] See Dutch Ministry of the Interior and Kingdom Relations, 'Impact Assessment. Fundamental Rights and Algorithms Impact Assessment' (2022) <https://www.government.nl/documents/reports/2021/07/31/impact-assessment-fundamental-rights-and-algorithms> accessed 12 February 2024.
[108] Ibid, 16-17.
[109] In this regard, for example, competing interests in the field of fundamental rights may justify the use of AI systems to promote some prevailing rights or freedoms without providing an explicit legal basis.
[110] Ibid, 13-15. In the case of multiple public values, the authors also introduce the idea of value weighting, which would require a specific methodological clarification, a given benchmark, and may be very challenging when considering a variety of socially relevant values, which may also change from one context to another.
[111] See Mantelero, Beyond Data (fn. 7), Ch. 3.
[112] For example, the How phase in Dutch Ministry of the Interior and Kingdom Relations, Impact Assessment (fn. 107), 47-65, includes elements relating to all the three components of the traditional risk assessment cycle, such as values and interests impacted, mitigation measures centered on human supervision and transparency, potential risks, auditing.
[113] Ibid, 25-27 and 33-46.
[114] Ibid, 28-29.
[115] Ibid, 30-32.
[116] See, e.g., ibid, 30-32 on data security issues, which may affect rightsholders in various way depending on the specific AI application and related interests involved, and may also have no impact on fundamental rights.
[117] See, e.g., European Environment Agency, 'Environmental issue report No 4. Environmental Risk Assessment - Approaches, Experiences and Information Sources' (1998), 90 <https://www.eea.europa.eu/publications/GH-07-97-595-EN-C2/riskindex.html> accessed 20 December 2023.

At the end of this first phase of the assessment, the FRIA should provide a first outline of the AI-related risk, starting from the needs analysis and the description of the system, then going on to consider the contextual scenario of fundamental rights (including the checks already in place) and the potentially impacted areas. With regard to the latter, it is important to focus on the impact on rights deriving from criticalities in the functioning of AI. In this regard, for example, the Algorithmic Impact Assessment (AIA) Tool developed by the Government of Canada[119] evaluates the impact that the adoption of an automated decision may have in general on the rights or freedoms of individuals. However, in FRIA, the focus is not on the general effects of AI-driven decisions (for example, an automated decision regarding free access to health services has in itself an impact on the right to health and its exercise), but on the impact of prejudicial decisions resulting from problems in the design, context, and functioning of the algorithm, such as denial of access to health services due to biases that create an illegitimate prejudice to this right.

Based on the analysis carried out in this section, best practices in developing this first phase of the FRIA should emphasise the relevance of the fundamental question of alternatives to AI, which could also be addressed using a SWOT (strengths, weaknesses, opportunities, and threats) analysis, and should consider the limitations that affect generic checklists, which should be combined with appropriate contextualisation provided by the experts carrying out the FRIA.[120] In designing a contextualised checklist inspiration can be drawn from the existing models, but remembering their limitations and not excluding other methods of analysis.

With a view to establishing common best practices for FRIA, existing checklists can be used as input for this phase,[121] but it is difficult to see the need to formalise them into a comprehensive and standardised set of questions.[122] A different conclusion can be drawn regarding the importance of outlining the key areas to be considered in the planning and scoping phase, both in general and for sector-specific AI applications. In this respect, some common elements relating to the characteristics of the AI system and its scope, as well as the socio-technical context of use, need to be considered in all evaluations in this first phase.

In the same way, some sector-specific issues – such as the nature and value of doctor-patient interaction in medicine, or the evolving nature of case law with respect to path-dependent AI systems to be used in the judicial domain – should be included amongst the key elements that characterise the dynamics of each sector under consideration. However, this task does not necessarily need to be delegated to the legislator, AI supervisory bodies or standard-setting bodies, as it can be carried out by the relevant stakeholders in each context, who can outline the key guiding principles in guidelines or similar tools to be used during the assessment practice when defining sectoral characteristics.

On the basis of the information gathered in the planning and scoping and risk identification phase, it is possible to carry out an in-depth analysis of the level of impact on individual rights, to quantify this impact and to prevent or reduce it through appropriate measures. The combination of these activities constitutes the impact assessment stage described in the next section.

### 5.2. Second FRIA block: risk analysis and management

Risk quantification is an essential part of any impact assessment. In this respect, Article 27 of the AI Act provides for less detail than the text proposed by the European Parliament. This will require further effort from AI operators in order to comply with the law. However, based on the observations made in Section 4, this phase should include (i) the analysis of the level of impact of the AI system on potentially affected fundamental rights; (ii) the identification of appropriate measures to prevent or mitigate[123] the risk, taking into account their impact on the risk level according to context-specific scenario analysis; (iii) the implementation of such measures, and (iv) the monitoring of the functioning of the AI system in order to revise the assessment and the adopted measures should technological, societal and contextual changes affect the level of risk or the effectiveness of the adopted measures.

Risk in terms of impact on fundamental rights can be represented on a scale from a minimum – assuming, in line with general risk theory, that there is no zero risk – to a maximum. The use of scaling is common in social research, but many of the phenomena that affect people in their social and relational lives are difficult to measure by using cardinal variables.[124] For example, we can use sensors to measure the temperature of a place by quantifying it with an exact number, but we cannot quantify in the same way the impact of a credit scoring system on some groups of people in terms of discrimination.

Adoption of a scale along a continuum (from a minimum to a maximum level of impact) makes it possible to use ordinal variables (e.g., low, medium, high, very high). The use of scaling and associated variables make it possible to compare different situations using the same variables, such as the various levels of impact on non-discrimination produced by a credit scoring system using a particular algorithm or introducing some modification to it. These ordinal variables can therefore be used for FRIA to 'measure' the impact on a range-based quantification of the risk (low, medium, high, very high).

---

[118] See, e.g., Algorithmic Impact Assessment model developed by the Government of Canada (fn. 101); The Alan Turing Institute, 'Human Rights, Democracy, and the Rule of Law Assurance Framework for AI Systems: A proposal prepared for the Council of Europe's Ad hoc Committee on Artificial Intelligence' (2021) 46-49 and 72-125 <https://rm.coe.int/huderaf-coe-final-1-2752-6741-5300-v-1/1680a3f688> accessed 12 February 2023, which provides a list of questions concerning the product/service, the context of use, and the relevant shareholders and affected groups, including vulnerability issues.
[119] See fn. 101.
[120] The role of experts in conducting the FRIA is crucial, not only for the quality of the overall results, but also because it mitigates some of the common shortcomings of using questionnaires for risk assessment, such as unintentional bias in question selection, respondent confirmation bias, and information gaps.
[121] While it is possible to define some standard general questions (e.g. on data quality and AI typology, and general scenario descriptions including potentially affected categories), it is up to the expert (or team of experts) carrying out the assessment to define the key relevant elements to be considered and investigated, based on domain knowledge. See also The Danish Institute for Human Rights, 'Human rights indicators for business' (2019) <https://www.humanrights.dk/tools/human-rights-indicators-business> accessed 07 April 2024.
[122] A standardised checklist would be relevant if the impact assessment were carried out by a layperson without the appropriate domain expertise, but this option contradicts the nature, objectives, quality and effectiveness of the impact assessment process, not only in AI but in general.

[123] In line with the approach taken in HRIA, risk mitigation refers only to measures that reduce the likelihood of the risk occurring (from high to medium or low), as human rights impact assessments do not accommodate residual impacts, where mitigating the consequences (severity) would actually mean accepting some residual prejudice to human rights. See also Radu Mares, 'Securing Human Rights through Risk-Management Methods: Breakthrough or Misalignment?' (2019) 32(3) Leiden Journal of International Law 517, 526-527, https://doi.org/10.1017/S0922156519000244. For a more restrictive interpretation, see Karen Yeung and Lee A. Bygrave, 'Demystifying the Modernized European Data Protection Regime: Cross-Disciplinary Insights from Legal and Regulatory Governance Scholarship' (2022) 16(1) Regulation & Governance 137–155, 146, https://doi.org/10.1111/rego.12401: with regard to the impact on fundamental rights in data protection, they conclude that processing operations that would clearly violate fundamental rights cannot lawfully proceed, while "proposed processing operations that could be regarded as "borderline" in that there is some uncertainty about whether they would, if implemented, constitute a violation of fundamental rights […] can fall within the scope of Article 35(1) if the severity and probability of the threatened "risk" to fundamental rights are "high". The impact assessment is therefore limited to the latter cases.
[124] Cardinal variables represent a real measurable value, such as the temperature in a room.

However, an abstract concept such as impact on rights needs to be operationalised in order to be assessed by using variables. It is therefore necessary to identify the risk components which, in terms of impact on rights, consist of two key dimensions: the likelihood of adverse impact and its severity.[125] The combination of the variables relating to these two dimensions provides a risk index that is assessed for each of the rights and freedoms potentially affected.[126]

In terms of social research methodology, it is also possible to create a composite index combining all potential impacts on rights, as well as a composite index combining the results of the FRIA and the results of the Conformity Assessment to create an overall impact index. However, this approach conflicts with the legal approach to fundamental rights where each right must be considered independently in terms of its potential prejudice. The fact that one right is less affected than another cannot lead to any form of compensation.[127]

In the field of law, the only way of settling different interests is through the balancing test in the presence of conflicting rights, but this test follows the assessment of the level of impact on each right. The balancing test does not relate to the level of risk for the affected rights, but to the prevalence of one interest over another. It should therefore be considered as an external factor, to be taken into account only after impact on individual rights has been assessed, and which may influence the results of the impact assessment by making acceptable a situation of high impact due to the presence of a prevailing competing interest.[128]

Based on these considerations, a FRIA model will define a risk index for each potentially impacted right using two dimensions: likelihood and severity.[129] As both dimensions are contextual and need to be 'measured', it is important to identify the key variables to be used for this purpose. In this context, likelihood is understood as a combination of (i) the probability of adverse outcomes and (ii) exposure. The first variable relates to the probability that adverse consequences of a given risk will occur and the second variable relates to the extent to which people potentially at risk could be affected. As far as exposure is concerned, it should be noted that the focus is on the group of potentially affected people and not on the whole population.[130]

The severity of the expected consequences is based on two variables: (i) the gravity of prejudice in the exercise of rights and freedoms (gravity),[131] including taking into account group-specific impact, vulnerability[132] and dependency situations, and (ii) the effort to overcome it and to reverse the adverse effects (effort).[133]

Both likelihood and severity need to be assessed on a contextual basis, and the involvement of relevant stakeholders can be of help.[134] As is common in risk assessment, the estimation of likelihood is based on both previous cases, looking at comparable situations, and the use of analytical and simulation techniques, based on possible scenarios of use. The same approaches are also used to estimate level of severity, but in this case with greater emphasis on legal analysis regarding the gravity of prejudice, which should be assessed with reference to the case law on fundamental rights and the legal framework.

Based on the values of likelihood and severity derived from the variables above, a risk index showing the overall impact is determined for each of the rights and freedoms considered,[135] a radial graph can be used to provide an overview of the various impacts generated by the AI system, to prioritise the areas for intervention, and to compare the overall effectiveness of different design/use options in reducing adverse impacts.

---

[125] See Article 3(2) of the AI Act, which states that " 'risk' means the combination of the probability of an occurrence of harm and the severity of that harm". In terms of the impact on fundamental rights, it seems difficult in the AI Act to have cases where the risk is dissociated from the harm, as is the case in the GDPR, where the violations of the Regulation in themselves reduce the protection of personal data. In the GDPR, this is due to the component of procedural approach that characterises data protection. The latter relies on safeguards in processing operations and therefore the mere lack of compliance with the GDPR obligations can reduce the level of protection of the right, regardless of any concrete harm; see, e.g., ECJ, BL v. MediaMarktSaturn Hagen-Iserlohn GmbH, C-687/21, 25 January 2024, paras 56-61, ECLI:EU:C:2024:72, and ECJ, Österreichische Post, C-300/21, 4 May 2023, EU:C:2023:370, paras 33-42. Furthermore, data protection is an enabling right in relation to other fundamental rights, which means that data processing can be the first stage of a potential subsequent harm due to a violation of another right from which the harm will arise. However, in regulating the FRIA in Article 27(1)(d), it would have been more accurate to refer to the notion of (adverse) impact, including any case of prejudice and limitation of fundamental rights, as at the beginning of Article 27, rather than harm. See also Office of the High Commissioner for Human Rights, 'The corporate responsibility to respect human rights. An Interpretive Guide' (New York - Geneva, 2012) <https://www.ohchr.org/sites/default/files/Documents/Publications/HR.PUB.12.2_En.pdf> accessed 14 November 2023, 15 ("An "adverse human rights impact" occurs when an action removes or reduces the ability of an individual to enjoy his or her human rights").

[126] For an example of this exercise, see Mantelero, Beyond Data (fn. 7), 60-76.

[127] For example, a discriminatory credit scoring system cannot be considered acceptable because the level of protection of personal data is high, with a trade-off between non-discrimination and data protection.

[128] See also below Section 5.2.1.

[129] See Article 3(2), AI Act. See also Article 34(1) DSA ("This risk assessment shall be specific to their services and proportionate to the systemic risks, taking into consideration their severity and probability"), and National Institute of Standards and Technology – NIST, Artificial Intelligence Risk Management Framework (fn. 16), 4, which defines risk as "the composite measure of an event's probability of occurring and the magnitude or degree of the consequences of the corresponding event".

[130] In the case of use of AI to provide national financial aid to families with 3 or more children based on various socio-economic parameters, for example, only the group of the families with these characteristics will be relevant in terms of exposure, not the entire population. Moreover, the extent of exposure may also be limited by contextual factors, such as in the case of an AI-based facial recognition system mainly trained using white Caucasian faces; the use of this system in a multiethnic context will affect non-Caucasian people, reducing accuracy of recognition, not the entire group of those who potentially interact with the system.

[131] The gravity/seriousness of prejudice to a human right is usually assessed according to the following three elements: (i) its intensity, (ii) the consequences of the violation, and (iii) its duration, where the intensity of the violation is related to the importance of the violated protected legal interest. See also Altwicker-Hamori et al., 'Measuring Violations of Human Rights: An Empirical Analysis of Awards in Respect of Non-Pecuniary Damage Under the European Convention on Human Rights' (2016) 76 Zeitschrift für ausländisches öffentliches Recht und Völkerrecht (ZaöRV)/Heidelberg Journal of International Law (HJIL) 1-51; European Court of Justice, 'Right to Respect for Private and Family Life, Home and Correspondence: A Practical Guide to the Article 8 Case-Law of the European Court of Human Rights». Council of Europe', 31 August 2022 <https://www.echr.coe.int/documents/d/echr/guide_art_8_eng> accessed 10 December 2023.

[132] On vulnerability and data use, see also Gianclaudio Malgieri. 2023. Vulnerability and data protection law (Oxford: Oxford university press).

[133] In this regard, in a different context of technology assessment of potential high-risk systems to be included in Annex III, Article 7(2)(i) AI Act states that "adverse impact on health, safety or fundamental rights, shall not be considered to be easily corrigible or reversible", a conclusion that seems too rigid, as there are cases where impacts on fundamental rights can be appropriately mitigated and reversed by contextual actions. See, for example, the case of inappropriate discriminatory language used by a companion robot in a classroom in relation to the teacher's intervention.

[134] See the following section.

[135] See the following section.

Factors that may exclude risk, from a legal perspective, such as the mandatory nature of certain impacting characteristics, should also be considered. After initial adoption of appropriate measures to counter the identified risks, further rounds of assessment can be conducted depending on the level of residual risk and its acceptability.

With regard to these methodological requirements,[136] based on risk assessment theory and the previous experience of its application to human rights, only a limited number of impact assessment models dealing with AI and fundamental rights have gone beyond the mere awareness-raising approach and introduced some elements of risk assessment and quantification.

An attempt to create a risk index for algorithmic impacts has been made by the Algorithmic Impact Assessment Tool developed by the Government of Canada, discussed above in relation to the planning and scoping phase. The Canadian model uses an impact index (Raw impact score) composed of six different indicators (Project, System, Algorithm, Decision, Impact, and Data), operationalised by a set of variables for each of them. In order to obtain the overall level of impact, another index (Mitigation score) based on two indicators (Consultation, De-risking and mitigation measures) is combined with the impact index.

The Canadian model gives limited importance to impact on fundamental rights, which, at most and in its broadest interpretation, can account for 20 points out of a maximum of 42 points for the Impact indicator and 126 points for the overall Raw impact index. The model gives more weight to data management (44 points) and significant weight to project design (27 points); in addition, both of these indicators include elements related to impact on fundamental rights, such as vulnerability, disability and data protection.

The Canadian model, as far as fundamental rights are concerned, suffers from a number of drawbacks. These include fragmentation of the relevant elements into different indicators, limited attention paid to fundamental rights, and the combination of different indicators in cumulative indicators that ignore the specificity of the impact on each individual right and the above-mentioned shortcomings in providing a single overall assessment.[137]

Moreover, the scoring of variables is not described methodologically and is unclear, with scores that appear to be contradictory.[138] In addition, different scoring ranges are used for different variables, which could be justified by the potential impact of the specific aspect on the overall risk, but the model does not provide transparent information on the criteria used for this differentiation, making it difficult to understand how the ranges of scores vary according to the different aspects considered.

The scoring is also unclear as regards the composition of the overall impact, which combines the impact (Raw impact score) and mitigation measures (Mitigation score) indices. According to the Canadian model "if the mitigation score is less than 80% of the maximum attainable mitigation score, the current score is equal to the raw impact score", whereas if the 80% threshold is exceeded, the Raw impact score is reduced by 15%. Both these thresholds (80% and 15%) are set without any evidence for their value.

Another example of previous experience that can be used to design the FRIA and stimulate reflection on best practice is the Impact Assessment Fundamental Rights and Algorithms (FRIAI) proposed by the Dutch Ministry of Interior and Kingdom Relations,[139] mentioned above in relation to the general methodological approach. The Dutch model focusses on two elements as regards the impact on fundamental rights: the seriousness of the infringement and the balancing test to justify restrictions on fundamental rights.[140]

Depending on the seriousness of interference with fundamental rights, various conditions for the use of AI are envisaged: justification for serious interference, due diligence for medium-serious interference, and no specific conditions for less serious interference.[141] The notion of seriousness used in this model refers to "to what extent a specific algorithm (or its application) will affect the core of a fundamental right",[142] but is only one of the dimensions (variables) to be considered in assessing the risk of impact on fundamental rights. Without considering at least the probability of negative impact on the core of fundamental rights concerned, the model is incomplete and unable to fully assess the level of risk.[143]

In addition, this model links the effectiveness of the algorithmic solution in achieving its objective with the seriousness of the impact on fundamental rights in terms of trade-offs.[144] This may be problematic from a fundamental rights perspective as effectiveness does not necessarily result in beneficial effects on competing rights that might justify a compression of the fundamental rights concerned.[145] In this respect, the Dutch model places fundamental rights and other "interests, objectives, and public values"[146] on the same level in a way that contradicts the level of protection afforded to fundamental rights.

---

[136] See fn. 97.

[137] The indicator concerning the Impact Assessment includes, inter alia, the five variables relating to fundamental rights measured on a four-level scale (little no impact, moderate impact, high impact, very high impact) and concerning the impact on "the rights or freedoms of individuals", "the equality, dignity, privacy, and autonomy of individuals", "the health and well-being of individuals", "the economic interests of individuals", "the ongoing sustainability of an environmental ecosystem", combining various rights and freedoms in a single variable.

[138] For example, in creating the Risk Profile indicator, the model gives +3 risk points for particularly vulnerable clients and +4 risk points for creation/exacerbation of barriers for people with disabilities. However, disability should be considered as a vulnerability, with the result that this scoring gives it too much weight compared to other vulnerabilities. In addition, it is unclear why other vulnerabilities lead to a lower increase in risk compared to disability. For the same indicator, one variable concerns the stakes of decisions and whether they are very high, but this is a rather open-ended question with a wide range of variability in the respondent's assessment.

[139] See Dutch Ministry of the Interior and Kingdom Relations, Impact Assessment (fn. 107)

[140] Ibid, 67.

[141] In terms of risk assessment, this means that when the impact is high only a prevailing competing interest may justify the use of AI systems, based on the balancing test.

[142] See Dutch Ministry of the Interior and Kingdom Relations, Impact Assessment (fn. 107), 75.

[143] In this respect, the model used both the notion of interference and that of risk, without considering the difference. See Katerina Demetzou, 'Data Protection Impact Assessment: A tool for accountability and the unclarified concept of 'high risk' in the General Data Protection Regulation' (2019) 35(6) Computer Law & Security Review: 105342. https://doi.org/10.1016/j.clsr.2019.105342 ("risk is a hypothetical event while an interference refers to an event that has already taken place. This means that interference is assessed and evaluated only in terms of severity, while the element of likelihood is of no relevance. Hence, the concept of 'interference' is useful in our analysis when it comes to evaluating the severity of a risk but not when it comes to its likelihood.").

[144] See, e.g., Dutch Ministry of the Interior and Kingdom Relations, Impact Assessment (fn. 107), 77 ("The red colour code (serious interference with a fundamental right) means that a far more rigorous investigation may be expected to establish whether the algorithm will result in achieving the set objectives than is the case for the green code") and 79 ("The question may come up what the trade-off should be if an alternative appears to be slightly less effective but does appear to affect the fundamental right less").

[145] For example, a rigorous investigation demonstrating the cost-reducing effects of real-time driver tracking on traffic management does not change the impact on the rights at risk and does not justify the solution in view of the proportionality and fundamental rights issues it raises.

[146] Ibid, 80.

Another model of impact assessment has been proposed by The Alan Turing Institute, as part of the work of the Council of Europe's Ad hoc Committee on Artificial Intelligence.[147] This HUDERIA (Human Rights, Democracy, and the Rule of Law Impact Assessment) model includes impact on democracy, which may be problematic from a methodological perspective, as it is difficult both to define a benchmark for level of democracy and to quantify the impact of an AI solution on it.

Regardless of this aspect, the fundamental rights part of the HUDERIA has several shortcomings that should be taken account of when designing a model for FRIA. As in the Canadian Algorithmic Impact Assessment Tool, different rights are grouped together according to a list of principles and priorities,[148] but the most relevant critical issue concerns the creation of the index.

The HUDERIA uses an index to assess the overall level of impact on fundamental rights, which combines severity and likelihood through summation.[149] Two variables are used for severity (Gravity potential and Number of affected rightsholders) with a maximum cumulative score of 6, and one variable is used for likelihood with maximum score of 4. The resulting index thus gives more weight to severity than likelihood, which the authors claim is "in accordance with its priority as a 'predominant factor' in human rights risk", but they do not provide any further arguments or clarification on how the ratio between the two variables was determined.

Regarding severity, its gravity component is based on a four-level scale that combines gravity of prejudice (catastrophic, critical, serious, and moderate/minor) and the temporal extent of the prejudice (irreversible, enduring, temporary, not enduring/temporary) in a fixed arrangement of pairs at the same level of the scale. Thus, for example, a catastrophic prejudice is always irreversible, and a critical prejudice is always enduring. This scale therefore not only includes in the variable described (gravity) another dependent variable (temporal extent), but also assumes a fixed correspondence between them, which is not the case in real life. For example, a serious prejudice to a fundamental right can be irreversible.

Other examples of potential issues in the definition of relevant variables are provided by the way in which the HUDERIA has defined the extent of impact, in terms of Number of affected rightsholders. In doing so, it uses ranges based on absolute numbers to estimate the extent of the impact in terms of people affected[150] and timescale,[151] thus missing the contextual dimension of impact assessment.[152] Similar problems in framing the variables are present in the likelihood levels, which include references to both the probability of the event and to the level of risk in general, although the latter relates to both severity and likelihood.[153]

A crucial aspect of risk assessment concerns the design of the risk matrix, where the range distribution used to construct it must be methodologically justified. Given the diversity of AI applications and use contexts, there is no one-size-fits-all risk distribution, but the matrix should be designed with context in mind. In this respect, fixed matrices based on predefined ranges and mathematical estimation, such as in the HUDERIA,[154] require a specific description of the criteria used to establish both the ranges and the matrix distribution.[155] Furthermore, matrices with a fixed distribution are unlikely to be consistent with all the potential application scenarios for AI.

Similar issues are also present in the model developed by UNESCO for Ethical Impact Assessment in AI, which has a different scope but where several variables are used without clarifying how they are combined in the overall impact.[156]

This empirical analysis of some proposed models reveals the need for extreme care, in operationalisation of the FRIA, as regards the creation of indices and indicators, the definition of the variables, and the design of the risk matrix. The fundamental rights impact assessment model must consider all major risk dimensions, carefully selecting relevant variables and their combination, and avoiding unjustified overweighting. The result must be consistent with the legal framework and theory of fundamental rights, identify the relevant rightsholders affected, exclude cumulative assessment of the impact on different rights and the fragmentation of impacted rights into different components, thus avoiding counting them twice, both as an individual element and as a component of the general right.

When using matrices to assess impacts, the relationship between the relevant risk components should be carefully considered in line with the fundamental rights framework; this suggests the need to avoid purely mathematical approaches to scaling, be transparent in the scaling criteria, and clearly define the relationship between the risk components.

All these technical aspects of operationalising the assessment of impacts on fundamental rights underline its necessarily expert-based nature. The use of variables also demonstrates, as in the case of the DPIA and HRIA, the dependence of the results on contextual factors that may change over time during the life cycle of AI systems.

### 5.2.1. The risk quantification matrix

The use of matrices to construct risk indices is recommended in many risk-based impact assessment models and standards, partly because they are relatively easy to use and explain. As a risk matrix is a graph that combines two dimensions using colours to reflect different levels of risk, they are useful for assessing indices generated by different variables. For this reason, as described in the previous section, they are used in the FRIA to define the level of impact on each right concerned.

In the light of the critical issues raised in the examination of the main models for the FRIA, the methodology proposed here uses a risk index for each potentially impacted right based on a matrix combining two dimensions (likelihood and severity). Each of these dimensions results from the combination of two pairs of variables, also constructed using matrices: the probability of adverse consequences, and exposure, for likelihood; the gravity of prejudice, and the effort to overcome it and to

---

[147] See The Alan Turing Institute, Human Rights, Democracy, and the Rule of Law Assurance Framework for AI Systems (fn. 118).
[148] Ibid, 49-53 and Tabel 1.
[149] Ibid, 63 and 208.
[150] Ibid, 83-84.
[151] Ibid, 85-86.
[152] Ibid, 66. In terms of people affected, the extent of impact needs to be calculated as a percentage of the total number of components of the target category; it is therefore not appropriate to create a scale with fixed ranges of rightsholders (1-10,000; 10,001-100,000; 100,001-1,000,000; over 1,000,000). For example, an AI system used for the national health service may discriminate against people affected by a particular rare disease that affects only 5,000 people in a county. If the proposed scale is adopted, the lower level will be assigned even though 100% of the target population is potentially affected. The same consideration can be expressed with regard to the proposed ranges of timescale (less than a year; 1-10 years; 10-20 years; 20-60 years; over 60 years), which are not justified and seem inconsistent with the rapid evolution and nature of most AI technologies. More generally, it seems difficult to define a common timescale in relation to affected groups, given that the impact of AI may vary according to the rights and persons affected.
[153] Ibid, 62 (see, e.g. the Likely level of the Likelihood levels which is described as "The risk of adverse impact is moderate; the harm is possible and may occur").
[154] Ibid, 67.
[155] For example, the HUDERIA matrix is polarised between low risk level and high/very high risk levels with a small area for medium risk. This is an unusual distribution that should be clarified, as should the ranges adopted for the scale. Moreover, the scale does not include some possible values (e.g., Severity scored 2), making it unclear how to define the matrix combination in these cases.
[156] See UNESCO. 2023. Ethical Impact Assessment (fn. 102).

reverse adverse effects, for severity.[157]

Some impact assessment proposals[158] also consider exposure in terms of duration over time. However, in the case of AI projects such a variable can be difficult to estimate, significantly increasing the granularity of the analysis and its reasonable feasibility.[159]

Although scaling and matrix design is not a particularly complicated exercise, its role in developing the relevant indicators and final index for decision-making in risk management means that it is crucial to keep high levels of accuracy and, in terms of accountability, to keep a record of all choices made in the matrix design. The same care must be taken with risk evaluation, to avoid inaccurate and incomplete risk analysis.[160]

As the various components of the impact on fundamental rights cannot be quantified in a precise and granular manner, variables, indicators and indices should not use numerical values. They should be elaborated in the FRIA by using qualitative risk matrices,[161] where each risk component varies on a scale along a continuum (from a minimum to a maximum level). In this way, scaling and ordinal variables (e.g., low, medium, high, very high) allow the FRIA to assess the overall impact on each right and how it may vary according to system design and deployment.[162]

There is no single risk matrix model to use in risk assessment. Practice in this field shows a variety of models, the most common of which are 3 × 3, 4 × 4, 5 × 5, 5 × 4 and 6 × 4 matrices, where the pairs of numbers indicate the number of ranges of the two scales defining the dimension under consideration. As the matrix refers to two independent variables, they can be evaluated according to scales that may differ in number of ranges, for example a 6 × 4 scale where six different ranges are provided for one variable and only four for the other.

Given the diversity of AI applications, the context of use, and the rights and rightsholders potentially affected, a tailored approach is required, where experts design the most appropriate matrix. This approach provides a high degree of flexibility: experts can fine-tune the model to the specific case and perform a detailed scenario-based analysis where they consider possible contextual scenarios and how the risk parameters can vary accordingly.

Based on the considerations expressed in previous work and the approach adopted there,[163] the 4 × 4 matrix may be the most appropriate in the context of HRIA/FRIA in AI, as it reduces the risk of average positioning, gives more attention to the high and very high levels in a way that is consistent with the focus on high risk in the current regulatory approach to AI, and does not excessively fragment the lower part of the scale, which is less relevant due to the aforementioned focus.[164]

In matrices, the use of cardinal numbers to express the risk ranges resulting from the contextual assessment should be reconsidered in favour of descriptive labels for the different combinations of levels in the colour scale,[165] as follows in this example of a severity matrix :[166]

| Severity Matrix | | Gravity | | | |
|---|---|---|---|---|---|
| | | Low | Medum | High | Very high |
| **Effort** | Low | Low/L | Low/Medium | Low/H | Low/VH |
| | Medium | Medium/L | Medium/M | Medium/H | Medium/VH |
| | High | High/L | High/M | High/H | High/VH |
| | Very high | Very high/L | Very high/M | Very high/H | Very high/VH |

| Severity levels |
|---|
| Low |
| Medium |
| High |
| Very high |

As concerns the possible measures to address the level of impact detected and to prevent or mitigate[167] it, their contextual nature – dependant on the AI system used, its characteristics and the context of its use – makes it difficult to provide a list of them. However, some general guidelines can be provided in this regard, with the caveat that they will be contextualised and expanded when the FRIA is carried out, according to the specificity of each case.

In this regard, for example, the Canadian Algorithmic Impact Assessment Tool emphasises the measures related to internal organisation and provides a useful list of possible general mitigation measures, as well as suggesting a series of questions that can be asked to monitor the effectiveness of such measures through auditing and accountability requirements.

In general terms, various solutions are used to manage risk: it can be prevented, reduced, or retained by those who created it, and transferred to other entities. From a legal point of view, the latter two options are *ex post* remedies, framed by tort law in terms of liability and insurance, or socialisation of the damage if it is covered by the state; they are therefore outside the scope of this analysis focused on the AI Act, which does not

---

[157] See Mantelero, Beyond Data (fn. 7), Ch 2. The use of this combination of parameters for risk matrices in AI-based cases was tested with positive feedback at the following expert meetings with human rights and AI experts, including representatives from industry, data protection authorities and government bodies: UNDP - Regional Workshop on Data Protection and The Impact of Technologies and AI on Human Rights in Eastern Europe and Central Asia, Istanbul, 27-28 October 2022; Catalan Data Protection Authority, Workshop for University Ethics Committees on Impact Assessment in AI, Barcelona, 12 June 2023; UNDP, Capacity Building for the Human Rights Impact Assessment (HRIA) in the development of technologies and AI systems - Human Rights and Equality Institution of Türkiye (TIHEK), Ankara, 1-2 February 2024.

[158] See, e.g., The Alan Turing Institute, Human Rights, Democracy, and the Rule of Law Assurance Framework for AI Systems (fn. 118), 85-86, but see above fn. 152 on the considerations on the inadequate variables used in defining the ranges of timescale.

[159] If we consider, for example, an AI system used for scoring in social aid, the potential impact on a given right (e.g., right to health) may have a different timescale according to the nature and extent of services provided/denied (e.g., a medical consultation rather than a surgical procedure) and the characteristics of the affected individuals (e.g., a young person or an older person). This wide variability makes it difficult to use timescale as a component of likelihood.

[160] See, e.g., Dutch Data Protection Authority - Department For The Coordination Of Algorithmic Oversight (DCA). 2024. AI & Algorithmic Risks Report Netherlands, fn. 42, 8, Graph 2, where the 'abuse of AI' is considered as impacting on disinformation only, while it may affect almost all fundamental rights due to the variety of AI uses. In the same line, negative effects of an 'inadequate design of systems and process' are not circumscribed to discrimination and privacy.

[161] Qualitative methods do not use numbers to express probability or impact but instead use established rating scales to describe probability and impact of risk. See above Section 5.2.

[162] See also Mantelero, Beyond Data (fn. 7), 54-59.

[163] Ibid.

[164] See, for example, a 6-grade-based scale for probability referring to Impossible, Improbable, Remote, Occasional, Probable, and Frequent options.

[165] The different colours in the matrix represent the level of severity, as shown in the severity table opposite.

[166] For an overview of the set of matrices necessary to fundamental rights impact assessment in AI, combining the different relevant variables, see Mantelero, Beyond Data (fn. 7), 54-59.

[167] See fn. 123.

include AI liability provisions.[168]

Regarding the former two options – that is preventing or reducing risk –, the measures to be adopted as a result of the impact assessment must prevent or mitigate[169] any risk to fundamental rights or, in the presence of prevailing interests according to the balancing test, they must prevent or mitigate any risk that is not justified by the pursuit of these interests, which act as exclusion factors with regard to residual risk.

In the FRIA, as well as in the fundamental rights session of the Conformity Assessment, the results of the initial assessment must include the level of impact, but also the exclusion factors, and the measures adopted. The exclusion factors refer to any elements that may justify the impact, such as a mandatory legal requirement and the balancing test with competing interests mentioned above. Finally, where specific measures are adopted, the assessment must demonstrate their effectiveness in reducing the initial level of risk.

*5.3. Third (underdeveloped) FRIA block: stakeholder participation*

Despite the importance of participation in improving risk awareness, reducing exclusion and biases, and better framing risks, including through co-design, the AI Act pays limited attention to this aspect.

While the text proposed by the European Parliament included some elements of participation in the FRIA,[170] the final version has not adopted a clear participatory approach, tush diverging from best practices in HRIA.[171] Only Recital 96 states that deployers "could involve" relevant stakeholders, including the representatives of groups of persons likely to be affected by the AI system, independent experts, and civil society organisations, leaving room for participation "when appropriate" and as a mere possibility ("could involve") "to collect relevant information necessary to perform the impact assessment".[172]

This limited reference to participation left two key questions unaddressed, in substantive and procedural terms respectively, concerning the representatives of groups of persons likely to be affected by the AI system: How can these groups be identified in the context of the operation of AI systems? What elements entitle an entity to be considered a representative? Here, the dynamic nature of AI classification and the lack of awareness of potentially affected individuals that are in the same situation may make this identification controversial.[173]

However, the absence of a specific participatory obligation in the AI Act does not mean that such an obligation cannot be required in the AI context on different legal grounds, including national law, and general human rights principles and obligations.

## 6. Framing the FRIA outside the AI Act perimeter

The AI Act concerns the use of a specific technology; in the absence of derogations, compliance with general principles and other existing legal obligations relevant in the context of AI is still required.[174] This is the case of fundamental rights, whose protection is an obligation for both AI providers and deployers as a result of the general protection afforded to these rights by international, EU and national laws, even in cases that fall outside the scope of the FRIA as defined in Article 27.

Hence, although the FRIA, under the AI Act, cannot be required outside the scope of Article 27, both providers and deployers can be sued for any violations of fundamental rights, and Member States can be held responsible for the lack of protection of human rights based on their international obligations. This is also relevant in the case of AI applications that are not considered high-risk under the AI Act,[175] but have an impact on fundamental rights, as the latter are protected regardless of the level of risk and how it is specifically considered in the AI Act.

Providers and deployers will therefore face the dilemma of whether to extend the mandatory FRIA, required by Article 27 only in certain cases, or to deal with potential litigation in the event of fundamental rights violations.[176] Indeed, the more effective *ex ante* protection provided by the AI Act does not exclude traditional remedies based on tort law and state liability for failure to protect fundamental rights.

*Ex post* remedies are less effective than preventing harm through risk management. However, AI operators are likely to be discouraged from viewing the absence of broad protection under the AI Act as a waiver of liability because of the cost of potentially massive multinational litigation, including to company reputation, and the interplay between fundamental rights and other important EU legislation providing for a wider range of remedies, such as blocking data processing and data erasure, under the GDPR, on orders from Supervisory Authorities.[177][178]

Furthermore, in relation to the interplay with the GDPR, it is important to highlight how the broad notion of personal data and the extensive use of AI in relation to individuals lead to the competence of the Supervisory Authorities to examine the functioning of the AI systems/models and their impact on fundamental rights under Article 35 of the GDPR.

Although Article 35 has been largely under-utilised in relation to the full range of fundamental rights,[179] the Supervisory Authorities can now use the full power of this provision to require a DPIA very close to the FRIA.[180] Article 35 also makes it possible to require this assessment not only in the high-risk cases identified in the AI Act. In addition, as Article 35 is accompanied by a specific sanction in case of non-compliance, this can fill the gap in the AI Act in terms of sanctions and contribute to a complementary role of the Supervisory Authorities in case national legislation and the competent authorities under the AI Act remain inactive.[181]

---

[168] See also European Commission, Proposal for a Directive of the European Parliament and of the Council on adapting non-contractual civil liability rules to artificial intelligence (AI Liability Directive), COM/2022/496 final, 2022.
[169] But on risk mitigation see fn. 123.
[170] See Article 29a(4), EP AI Act, in the text adopted by the European Parliament ("In the course of the impact assessment, the deployer, with the exception of SMEs, shall notify national supervisory authority and relevant stakeholders and shall, to best extent possible, involve representatives of the persons or groups of persons that are likely to be affected by the high risk AI system, as identified in paragraph 1, including but not limited to: equality bodies, consumer protection agencies, social partners and data protection agencies, with a view to receiving input into the impact assessment").
[171] See also United Nations - Human Rights Council. 2011. Guiding Principles on Business and Human Rights (fn. 17), n. 18. For a methodological approach to participation in AI design and impact assessment, see ECNL and SocietyInside, 'Framework for meaningful engagement' (2023) <https://ecnl.org/publications/framework-meaningful-engagement-human-rights-impact-assessments-ai> accessed 23 November 2023. See also Data & Society and ECNL, 'Recommendations for Assessing AI Impacts to Human Rights, Democracy, and the Rule of Law' (2021) <https://ecnl.org/sites/default/files/2021-11/HUDERIA%20paper%20ECNL%20and%20DataSociety.pdf> accessed 28 June 2024.
[172] See Rec. 96, AI Act ("Where appropriate, to collect relevant information necessary to perform the impact assessment, deployers of high-risk AI system, in particular when AI systems are used in the public sector, could involve relevant stakeholders, including the representatives of groups of persons likely to be affected by the AI system, independent experts, and civil society organisations in conducting such impact assessments and designing measures to be taken in the case of materialisation of the risks"). See also Rec. 65 on Conformity Assessment ("When identifying the most appropriate risk management measures, the provider should document and explain the choices made and, when relevant, involve experts and external stakeholders").

[173] See Alessandro Mantelero, 'AI and Big Data: A blueprint for a human rights, social and ethical impact assessment' (2018) 34(4) Computer Law & Security Review 754, 763-764, https://doi.org/10.1016/j.clsr.2018.05.017.
[174] See also Rec. 63 AI Act. See also Rec. 9.
[175] The advantage of fundamental rights impact assessment models lies in their ability to face the new development of AI, while regulatory approaches based on predefined high-risk or prohibited categories, such as the AI Act, lead to uncertain interpretations and the need to update the scenarios considered. See also Michael Pizzi et al. 'AI for Humanitarian Action: Human Rights and Ethics' (2020) 102(913) International Review of the Red Cross 145, 175, https://doi.org/10.1017/S1816383121000011.
[176] See also Rec. 170, AI Act.
[177] See fns. 4 and 10. See also ECJ, Újpesti Polgármesteri Hivatal, Case C-46/23, 14 March 2024, ECLI:EU:C:2024:239.
[178] See also in the same vein the observations on participatory obligations in Section 5.3 above.

Finally, the FRIA can also interact with the system of prohibitions established by the AI Act, with regard to both the exceptions included in Article 5 and the cases not listed therein. In relation to the former, the FRIA can highlight disproportionate and unnecessary violations of fundamental rights in some possible concrete applications of these exceptions. This will not simply block the use of the specific AI application, but will also open the door to a broader intervention by the ECJ,[182] providing a restrictive interpretation of the cases listed in Article 5, or the EU legislator, amending the AI Act.

As for unacceptable prejudice to fundamental rights arising from certain uses of AI not listed in Article 5, in the absence of an explicit derogation to the protection of these rights in the AI Act based on a balancing of interests, national courts can prohibit these uses even though not covered by the AI Act, as the latter does not provide an exhaustive list of possible insurmountable conflicts with fundamental rights. These rights remain protected by the EU Charters in any case where the law or the balancing test does not justify proportionate and necessary restrictions.

## 7. Conclusions

The main contribution of this article to the growing literature on the AI Act concerns two key elements of the FRIA, as framed in the AI Act and relevant in the field of AI systems: (i) the interpretation of Article 27 in a way that is consistent with the text, its implementation, and the context of fundamental rights protection and impact assessment at EU and global level; (ii) the operationalisation of the FRIA as defined in Article 27 in line with the methodological approach used in impact assessment in general and with specific reference to the area of rights and freedoms.

In examining the legal obligations set out in the AI Act relating to FRIA and their implementation, this article aims to meet the urgent needs of AI operators, EU and national supervisory bodies and regulators in defining one of the core elements of the EU regulation on AI, in terms of the importance of impacted interests and the increasing number of cases of concrete prejudice.

In addition, the proposed methodological approach for fundamental rights impact assessment[183] can be used beyond the limitations of the FRIA as framed by Article 27 and can be part of future methodological standards developed to operationalise the Conformity Assessment of the AI Act as regards fundamental rights.

Finally, by combining awareness-raising elements and risk quantification, the proposed way of implementing the FRIA creates an assessment tool that can foster both accountability and transparency, as well as support AI operators in developing and deploying their systems in a truly trustworthy and human-centric way.[184]

### Declaration of competing interest

The author declares that he has no known competing financial interests or personal relationships that could appear to have influenced the work reported in this paper.

### Data availability

No data was used for the research described in the article.

---

[179] See above Section 2.
[180] See also Article 27(4), AI Act.
[181] See Article 83(4), GPDR. See also Article 27(4), AI Act. See above Section 4.
[182] See also Recs 2 and 6, AI Act.
[183] See also fn. 97.
[184] See Rec. 1, AI Act.